\documentclass[11pt,ams]{article}
\usepackage{geometry}
\usepackage{dsfont}
\usepackage{amsmath}
\usepackage{amsfonts}
\usepackage{amssymb}
\usepackage{graphicx}%
\usepackage{subfig}
\usepackage{float}

\newcommand{\p}{\partial}

\newcommand{\s}{\sigma}

\renewcommand{\d}{\delta}
\renewcommand{\S}{\Sigma}

\newcommand{\e}{\varepsilon}

\begin{document}

\date{}
\title{\textbf{A study of the lightest glueball states  in $SU(2)$ Euclidean Yang-Mills
theory in the maximal Abelian gauge}}

\author{\textbf{M.~A.~L.~Capri}$^{a}$\thanks{capri@ufrrj.br}\,\,,
\textbf{A.~J.~G\'{o}mez}$^{b}$\thanks{ajgomez@uerj.br}\,\,,
\textbf{M.~S.~Guimaraes}$^{b}$\thanks{msguimaraes@uerj.br}\,\,,
\textbf{V.~E.~R.~Lemes}$^{b}$\thanks{vitor@dft.if.uerj.br}\,\,,\\
\textbf{S.~P.~Sorella}$^{b}$\thanks{sorella@uerj.br}\ \thanks{Work supported by
FAPERJ, Funda{\c{c}}{\~{a}}o de Amparo {\`{a}} Pesquisa do Estado do Rio de
Janeiro, under the program \textit{Cientista do Nosso Estado}, E-26/101.578/2010.}\,\,,\,\,\textbf{D.~G.~Tedesco}$^{b}$\thanks{dgtedesco@uerj.br}\,\,\\[2mm]
\textit{{\small $^{a}$ UFRRJ $-$ Universidade Federal Rural do Rio de Janeiro}}\\
\textit{{\small Departamento de F\'{\i}sica $-$ Grupo de F\'{\i}sica Te\'{o}rica e Matem\'{a}tica F\'{\i}sica}}\\
\textit{{\small BR 465-07, 23890-971, Serop\'edica, RJ, Brasil.}}\\
\textit{{\small {$^{b}$ UERJ $-$ Universidade do Estado do Rio de
Janeiro}}}\\\textit{{\small {Instituto de F\'{\i}sica $-$
Departamento de F\'{\i}sica Te\'{o}rica}}}\\\textit{{\small {Rua
S{\~a}o Francisco Xavier 524, 20550-013 Maracan{\~a}, Rio de
Janeiro, RJ, Brasil.}}}}
\maketitle
\begin{abstract}
A qualitative study of the lightest glueball states in Euclidean $SU(2)$ Yang-Mills theory quantized in the maximal Abelian gauge is presented. The analysis is done by generalizing to the maximal Abelian gauge the so-called replica model, already successfully introduced in the Landau gauge. As it will be shown, the gluon and ghost propagators obtained from the replica model are of the same type of those already introduced in \cite{Capri:2008ak}, whose  behavior turns out to be in agreement with that available from the lattice data on the maximal Abelian gauge.
The model turns out to be renormalizable to all orders, while enabling us to introduce gauge invariant composite operators for the study of the lightest glueballs $J^{PC}=0^{++}$, $2^{++}$ and $0^{-+}$. The spectral representation for the correlation functions of these operators are evaluated to the first order, and the corresponding spectral densities are shown to be positive. Under the assumption of Abelian dominance, it turns out that the hierarchy for the masses of the lightest glueballs in the maximal Abelian gauge is in agreement with that already obtained in the  Landau gauge, a feature which provides evidence for the gauge independence of the spectrum of the theory.
\end{abstract}
\baselineskip=13pt

\newpage

%%%%%%%%%%%%%%%%%%%%%%
\section{Introduction}
%%%%%%%%%%%%%%%%%%%%%%
The maximal Abelian gauge \cite{'tHooft:1981ht,Kronfeld:1987vd,Kronfeld:1987ri}  is extensively employed in the study of nonperturbative aspects of Yang-Mills theories. This is the case, for example, of the dual superconductivity mechanism for confinement \cite{Nambu:1974zg,Mandelstam:1974pi,'tHooft:1981ht}, according to which the low energy region of Yang-Mills theories should be described by an effective Abelian theory in the presence of monopoles. A dual Meissner effect arising as a consequence of the condensation of these magnetic charges might give rise to the formation of flux tubes which confine quarks. \\\\The possibility of achieving a description of the low energy regime of Yang-Mills theories in terms of an effective Abelian theory, also referred to as the Abelian dominance hypothesis \cite{Ezawa:1982bf,Suzuki:1989gp,Hioki:1991ai}, is being investigated since several years, from both theoretical and lattice simulations points of view,  see also \cite{Bornyakov:2011eq, Bornyakov:2011dj}) for more recent investigations. In the maximal Abelian gauge, the Abelian degrees of freedom are identified with the diagonal components of the gauge field corresponding to the diagonal generators of the gauge group. The remaining off-diagonal components decouple at low energies due to a dynamical generation of an off-diagonal mass, a feature which has already received analytical   \cite{Schaden:1999ew,Kondo:2001tm,Dudal:2004rx} and numerical evidence  \cite{Amemiya:1998jz,Bornyakov:2003ee,Mendes:2006kc}. \\\\In this work we pursue the investigation of the nonperturbative aspects of Yang-Mills theories in the maximal Abelian gauge by addressing the issue of the glueball spectrum and  of its comparison with the results already obtained in the case of the Landau gauge \cite{Capri:2010pg,Dudal:2010cd}. This goal will be achieved by combining several ingredients. Firstly, we shall generalize to the case of the maximal Abelian gauge the so called replica model, already introduced in the Landau gauge \cite{Sorella:2010it}. As we shall see, the replica model exhibits two nice features which are particularly useful in order to analyse the spectrum of the lightest glueballs. The first property concerns the gluon and ghost propagators, which turn out to be of the same kind of those already introduced in \cite{Capri:2008ak} and which are in agreement with the available lattice data  \cite{Mendes:2006kc}. Moreover, unlike the model introduced in  \cite{Capri:2008ak}, the replica model enables us to make use of genuine local gauge invariant composite operators whose correlation functions can be employed to study the glueball spectrum. In addition, and in complete analogy with the case of the Landau gauge, the replica model of the maximal Abelian gauge will be proven to be renormalizable to all orders. As such, the model displays useful nonperturbative features.  Secondly, we shall establish a  suitable mass formula for the glueballs by relying on a kind of inspired SVZ sum rules, already employed in the case of the Landau gauge  \cite{Capri:2010pg}. The combination of these two ingredients will enable us to provide a qualitative analysis of the location of the masses of the lightest glueball states. \\\\Having thus at our disposal a model which gives rise to  diagonal and off diagonal gluon propagators in agreement with those observed in lattice simulations, we employ it in order to study the correlation functions of the local composite gauge invariant operators with the quantum numbers of the three lightest glueballs, namely  $J^{PC}=0^{++}, 0^{-+}, 2^{++}$. In particular, thanks to the analytic structure of the diagonal gluon propagator, the correlation functions of the composite operators corresponding to the states $J^{PC}=0^{++}, 0^{-+}, 2^{++}$, can be cast in the form of a K\"all\'{e}n-Lehmann spectral representation with positive spectral densities, enabling us to provide a qualitative study of the mass spectrum of the lightest glueball states in the maximal Abelian gauge. It turns out that, under the assumption of the Abelian dominance \cite{Ezawa:1982bf,Suzuki:1989gp,Hioki:1991ai}, the resulting glueball spectrum is in agreement with that already obtained in the case of the Landau gauge \cite{Capri:2010pg,Dudal:2010cd}. This result can be taken as evidence of the nonperturbative gauge independence of the physical spectrum of the theory. \\\\The paper is organized as follows. In Sect.2 we give a brief account of the available  lattice numerical simulations on the gluon propagator in the maximal Abelian gauge. In Sect.3 we construct the replica model for the maximal Abelian gauge and we  show that the resulting gluon and ghost propagators have the same qualitative behavior of those reported in lattice simulations. In Sect.4 we introduce the gauge invariant composite operators corresponding to the three lightest glueball states $J^{PC}=0^{++}, 0^{-+}, 2^{++}$, and we determine the K\"all\'{e}n-Lehmann spectral representation for the corresponding correlation functions. Sect.5 is devoted to present our results about the mass spectrum of the states $J^{PC}=0^{++}, 0^{-+}, 2^{++}$. In Sect.6  we collect our conclusion. All technical details about the proof of the  renormalizability of the replica model in the maximal Abelian gauge can be found in the Appendix A, while in the Appendix B the plots of the mass hierarchy  of the states $J^{PC}=0^{++}, 0^{-+}, 2^{++}$ are displayed.
%%%%%%%%%%%%%%%%%%%%%%%%%%%%%%%%%%%%
\section{A look at the lattice data on the gluon propagator in the maximal Abelian gauge}
%%%%%%%%%%%%%%%%%%%%%%%%%%%%%%%%%%%%
Before giving a short account of the lattice investigations, let us remind that, in the $SU(2)$ maximal Abelian gauge, the gauge field ${\cal A}_{\mu}$ is decomposed into diagonal and off-diagonal components, according to
\begin{equation}
{\cal A}_\mu  = A^a_\mu\left( \frac{\sigma^a}{2} \right) + A_\mu  \left( \frac{\sigma^3}{2} \right)  \;, \qquad   a=1,2   \label{dec}
\end{equation}
where $\left( {\sigma^a}, {\sigma^3} \right)$ stand for the Pauli matrices. The diagonal component, corresponding to the diagonal generator ${\sigma^3}$, is $A_\mu$, while $A^a_\mu$, $a=1,2$, denote the off-diagonal components. In the maximal Abelian gauge, the fields $(A^a_\mu, A_\mu)$ are constrained by the following conditions \cite{'tHooft:1981ht,Kronfeld:1987vd,Kronfeld:1987ri}
\begin{eqnarray}
D^{ab}_{\mu}A^b_\mu&  = & \partial_{\mu}A^a_\mu -g\e^{ab}A_{\mu} A^b_\mu = 0  \;, \nonumber \\
\partial_\mu A_\mu & = & 0  \;. \label{mcond}
\end{eqnarray}
It is useful to remind here that the non-linear maximal Abelian gauge condition, $D^{ab}_{\mu}A^b_\mu=0$, can be implemented by minimizing  the auxiliary functional  \cite{Capri:2008ak}
\begin{equation}
{\cal F}[A] = \int d^4x A^a_\mu A^a_\mu  \;, \label{fa}
\end{equation}
with respect to the gauge transformations. The functional \eqref{fa}  possesses a lattice version \cite{Bornyakov:2003ee, Mendes:2006kc}., making possible the numerical implementation of the maximal Abelian gauge  by selecting the configurations which correspond to local minima of the auxiliary functional. \\\\The infrared behavior of the gluon propagator in the $SU(2)$ maximal Abelian gauge has been investigated by several authors through lattice numerical simulations. In \cite{Amemiya:1998jz}, both the diagonal and off-diagonal propagators were analysed in configuration space, exhibiting a behavior in agreement with the Abelian dominance. The off-diagonal gluon propagator turned out in fact to be suppressed with respect to the diagonal propagator. Later on, in \cite{Bornyakov:2003ee}, the diagonal and off-diagonal gluon propagators have been studied in momentum space. Also here, the off-diagonal propagator is suppressed with respect to the diagonal one. \\\\More recently, this issue has been addressed by the authors \cite{Mendes:2006kc} on a larger lattice. Both  gluon and ghost propagators have been analysed, and particular attention has been devoted to their behavior at very low momenta $p\approx 0$. According to \cite{Mendes:2006kc}, the transverse diagonal propagator can be fitted by
\begin{eqnarray}
\langle A_{\mu}(p) A_\nu(-p) \rangle  & = & \left( \delta_{\mu\nu} -\frac{p_\mu p_\nu}{p^2} \right) D_{diag}(p^2) \nonumber \;, \\
D_{diag}(p^2)& = & \frac{1+d p^2}{a+bp^2+cp^4} \;, \label{latt-diag}
\end{eqnarray}
were the parameter $m_{diag}=\sqrt{\frac{a}{b}}$ is found to be $m_{diag}\approx 0.72GeV$. It is worth to point out that the diagonal form factor $D_{diag}(p^2)$ does not vanish at zero momentum, $p=0$, and that expression \eqref{latt-diag} is precisely of the  same kind of that found in the case of the Landau gauge within the so called Refined Gribov-Zwanziger framework  \cite{Dudal:2007cw,Dudal:2008sp,Dudal:2011gd}  as well as within the Landau replica model \cite{Sorella:2010it}.  It is worth underlining that the nonvanishing of the gluon propagator at zero momentum in the Landau gauge has been also seen in lattice simulations, see for example  \cite{Cucchieri:2007rg,Bornyakov:2008yx,Dudal:2010tf,Cucchieri:2011ig}. As we shall see in the following, this is a very welcome feature which will play an important role in the discussion of the glueball spectrum in the maximal Abelian gauge.\\\\The best fit for the transverse off-diagonal gluon propagator is of the Yukawa type \cite{Mendes:2006kc}, namely
\begin{eqnarray}
\langle A^a_{\mu}(p) A^b_\nu(-p) \rangle_{transv.}  & = &\delta^{ab} \left( \delta_{\mu\nu} -\frac{p_\mu p_\nu}{p^2} \right) D_{off}(p^2) \;, \nonumber \\
D_{off}(p^2) & = & \frac{1}{a+bp^2} \;, \label{offd}
\end{eqnarray}
where the parameter $m_{off}=\sqrt{\frac{a}{b}}$ takes now the value $m_{off}=\approx 0.97GeV$. Although, as
expected from the hypothesis of Abelian dominance, the value of the off-diagonal parameter $m_{off}$ is greater than  $m_{diag}$, a clear evidence for the Abelian dominance would demand a more complete analysis, perhaps using larger lattices.    \\\\Although not needed for the purposes of the present investigation, it is worth to mention that the ghost propagator decomposes into  symmetric and anti-symmetric components in color space, {\it i.e.}
\begin{equation}
\langle {\bar c}^a(p) c^b(-p) \rangle = \delta^{ab} D_{symm} + \e^{ab} D_{asymm} \;. \label{gat}
\end{equation}
It is interesting to observe that the best fit for the symmetric form factor $D_{symm}$  obtained in \cite{Mendes:2006kc} is also of the Refined Gribov-Zwanziger type for the maximal Abelian gauge, as obtained in  \cite{Capri:2008ak}, namely
\begin{equation}
D_{symm} = \frac{1+d p^2}{a+bp^2+cp^4} \;, \label{ghsymm}
\end{equation}
with $a=0.45(1)GeV^2, b=1.1(3), c=0.73(30)GeV^{-2}, d=2.1(9)GeV^{-2}$. In particular, there is no sign of infrared enhancement in expression \eqref{ghsymm},  which attains a finite non-vanishing value at $p\approx0$. Concerning the antisymmetric component, $D_{asymm}$, the authors \cite{Mendes:2006kc} have provided a first preliminary evidence that it might be non-vanishing, a fact that could be linked to the existence of certain  non-vanishing ghost condensates of dimension two which can be naturally introduced in the maximal Abelian gauge, as argued in \cite{Capri:2007hw}. For specific details on the potential errors related to finite size and discretization effects we remind to the original lattice references \cite{Amemiya:1998jz, Bornyakov:2003ee,Mendes:2006kc}.

%%%%%%%%%%%%%%%%%%%
\section{Construction of the replica model in the maximal Abelian gauge}

%%%%%%%%%%%%%%%%%%%
%%%%%%%%%%%%%%%%%%%%%%%%%%%%%%%%%%%%%%%%%%%%%%%%%%%%%
%\subsection{Introducing the replica model in the MAG}
%%%%%%%%%%%%%%%%%%%%%%%%%%%%%%%%%%%%%%%%%%%%%%%%%%%%%
In order to construct the replica model, let us start with the Faddeev-Popov action for $SU(2)$ Yang-Mills quantized in the maximal Abelian gauge, namely
\begin{eqnarray}
\label{model}
S[A]&=&\int d^{4}x\,\biggl(\frac{1}{4}\,F^{a}_{\mu\nu}F^{a}_{\mu\nu}
+\frac{1}{4}\,F_{\mu\nu}F_{\mu\nu}
+ib^{a}\,D^{ab}_{\mu}A^{b}_{\mu}
-\bar{c}^{a}\mathcal{M}^{ab}c^{b}
+g\e^{ab}\bar{c}^{a}(D^{bc}_{\mu}A^{c}_{\mu})c\nonumber\\
&&\;\;\;\;\;\; +ib\,\partial_{\mu}A_{\mu}
+\bar{c}\,\partial_{\mu}(\partial_{\mu}c+g\varepsilon^{ab}A^{a}_{\mu}c^{b})\biggl)\,,
\label{action_S}
\end{eqnarray}
where $(F^{a}_{\mu\nu}, F_{\mu\nu})$ are the off-diagonal  and diagonal components of the field strength
\begin{equation}
F^{a}_{\mu\nu}=D^{ab}_{\mu}A^{b}_{\nu}
-D^{ab}_{\nu}A^{b}_{\mu}\,,\qquad
F_{\mu\nu}=\partial_{\mu}A_{\nu}
-\partial_{\nu}A_{\mu}
+g\varepsilon^{ab}A^{a}_{\mu}A^{b}_{\nu}\,,  \label{fab}
\end{equation}
and  $D^{ab}_{\mu}$ is the covariant derivative respect to the diagonal component, eq.\eqref{mcond},
\begin{equation}
D^{ab}_{\mu}=\d^{ab}\partial_{\mu}-g\e^{ab}A_{\mu}\,. \label{Dab}
\end{equation}
The indices $a,b,c$ appearing in eqs.\eqref{action_S}, \eqref{fab}, \eqref{Dab} refer to the off-diagonal components, and run from 1 to 2,  {\it i.e.} $a,b,c=1,2$. The fields $(b^a,b)$ are the Lagrange multipliers implementing the gauge conditions \eqref{mcond}, and $({\bar c}^a,c^a)$, $({\bar c},c)$ are the off-diagonal and diagonal Faddeev-Popov ghosts. Also, $\mathcal{M}^{ab}$ denotes the Faddeev-Popov operator of the maximal Abelian gauge, being given by
\begin{equation}
{\mathcal{M}}^{ab}=-D^{ac}_{\mu}D^{cb}_{\mu}
-g^{2}\varepsilon^{ac}\varepsilon^{bd}A^{c}_{\mu}A^{d}_{\mu}\,.  \label{fpm}
\end{equation}
The action \eqref{action_S} is left invariant by the nilpotent BRST transformations:
\begin{eqnarray}
&sA^{a}_{\mu}=-(D^{ab}_{\mu}c^{b}+g\e^{ab}A^{b}_{\mu}c)\,,\qquad
sA_{\mu}=-(\partial_{\mu}c+g\e^{ab}A^{a}_{\mu}c^{b})\,,&\nonumber\\
&sc^{a}=g\e^{ab}c^{b}c\,,\qquad
sc=\displaystyle\frac{g}{2}\e^{ab}c^{a}c^{b}\,,&\nonumber\\
&s\bar{c}^{a}=ib^{a}\,,\qquad s\bar{c}=ib\,,&\nonumber\\
&sb^{a}=0\,,\qquad sb=0\,,&
\end{eqnarray}
with
\begin{equation}
s S[A] = 0 \;. \label{invbrst}
\end{equation}
Furthermore, it is known that, for renormalization purposes \cite{Min:1985bx,Fazio:2001rm}, an additional BRST exact term proportional to a gauge parameter $\alpha$ has to be added to the action $S[A]$, which is replaced by
\begin{equation}
S_0[A]=S[A]+S_{\alpha}\,,\label{zero_action}
\end{equation}
where
\begin{eqnarray}
S_{\alpha} & = & s \left( \frac{\alpha}{2} \int d^4x (-i {\bar c}^a b^a + g \varepsilon^{ab} {\bar c}^a {\bar c}^b ) \right) \nonumber \\
& = & \frac{\alpha}{2}\int d^{4}x\,\left(b^{a}b^{a}
+2ig\varepsilon^{ab}\,b^{a}\bar{c}^{b}c
+g^{2}\bar{c}^{a}\bar{c}^{b}c^{a}c^{b}\right)\,.
\end{eqnarray}
The action $S_0[A]$ enjoys multiplicative renormalizability to all orders  \cite{Fazio:2001rm}. The limit $\alpha \rightarrow 0$ has to be taken after the removal of the ultraviolet divergences. \\\\We are now ready to start the construction of the replica model in the maximal Abelian gauge. To that end, we shall follow the procedure already outlined in the case of the Landau gauge \cite{Sorella:2010it,Capri:2010pg}. We first introduce a replica of the action $S_0[A]$, namely
\begin{eqnarray}
S_0[B]&=&\int d^{4}x\,\biggl\{\frac{1}{4}\,G^{a}_{\mu\nu}G^{a}_{\mu\nu}
+\frac{1}{4}\,G_{\mu\nu}G_{\mu\nu}
+i\bar{b}^{a}\,\bar{D}^{ab}_{\mu}B^{b}_{\mu}
-\bar{\omega}^{a}\overline{\mathcal{M}}^{ab}\omega^{b}
+g\e^{ab}\bar{\omega}^{a}(\bar{D}^{bc}_{\mu}B^{c}_{\mu})\omega\nonumber\\
&&+i\bar{b}\,\partial_{\mu}B_{\mu}
+\bar{\omega}\,\partial_{\mu}(\partial_{\mu}\omega+g\varepsilon^{ab}B^{a}_{\mu}\omega^{b})
+\frac{\alpha}{2}\left(\bar{b}^{a}\bar{b}^{a}
+2ig\varepsilon^{ab}\,\bar{b}^{a}\bar{\omega}^{b}\omega
+g^{2}\bar{\omega}^{a}\bar{\omega}^{b}\omega^{a}\omega^{b}\right)\biggr\}\,.
\label{action_zero_replica}
\end{eqnarray}
with
\begin{eqnarray}
G^{a}_{\mu\nu}&=&\bar{D}^{ab}_{\mu}B^{b}_{\nu}
-\bar{D}^{ab}_{\nu}B^{b}_{\mu}\,,\nonumber\\
G_{\mu\nu}&=&\partial_{\mu}B_{\nu}
-\partial_{\nu}B_{\mu}
+g\varepsilon^{ab}B^{a}_{\mu}B^{b}_{\nu}\,,\nonumber\\
\bar{D}^{ab}_{\mu}&=&\delta^{ab}\partial_{\mu}-g\varepsilon^{ab}B_{\mu}\,,\nonumber\\
\overline{\mathcal{M}}^{ab}&=&-\bar{D}^{ac}_{\mu}\bar{D}^{cb}_{\mu}
-g^{2}\varepsilon^{ac}\varepsilon^{bd}B^{c}_{\mu}B^{d}_{\mu}\,.  \label{fpmo}
\end{eqnarray}
The fields $(B_\mu, B^a_\mu, {\bar \omega}, \omega, {\bar \omega}^a, \omega^a, {\bar b}, {\bar b}^a)$ are a replica of those of expression \eqref{zero_action}. The action $S_0[B]$ is left invariant by the BRST symmetry
\begin{eqnarray}
&sB^{a}_{\mu}=-(\bar{D}^{ab}_{\mu}\omega^{b}+g\e^{ab}B^{b}_{\mu}\omega)\,,\qquad
sB_{\mu}=-(\partial_{\mu}\omega+g\e^{ab}B^{a}_{\mu}\omega^{b})\,,&\nonumber\\
&s\omega^{a}=g\e^{ab}\omega^{b}\omega\,,\qquad
s\omega=\displaystyle\frac{g}{2}\e^{ab}\omega^{a}\omega^{b}\,,&\nonumber\\
&s\bar{\omega}^{a}=i\bar{b}^{a}\,,\qquad s\bar{\omega}=i\bar{b}\,,&\nonumber\\
&sb^{a}=0\,,\qquad sb=0\,,&
\end{eqnarray}
\begin{equation}
s S_0[B] = 0\;. \label{brst-in-SU}
\end{equation}
Further, following the case of the Landau gauge \cite{Sorella:2010it,Capri:2010pg}, the two expressions $S_0[A]$ and  $S_0[B]$ are coupled through a soft term
\begin{equation}
S_{\vartheta}=i\sqrt{2}\vartheta^{2}\int d^{4}x\,A_{\mu}B_{\mu}\;, \label{softterm}
\end{equation}
where $\vartheta$ stands for a a free mass parameter. Therefore, we consider the action
\begin{equation}
{\widetilde S}=S_0[A]+S_0[B]+S_{\vartheta}\;. \label{actb}
\end{equation}
It is worth to point out that ${\widetilde S}$ enjoys the following discrete symmetry which, for obvious reasons, will be called  the mirror symmetry:
\begin{eqnarray}
(A^{a}_{\mu},A_{\mu})&\leftrightarrow& (B^{a}_{\mu},B_{\mu})\,,\nonumber\\
(c^{a},c)&\leftrightarrow& (\omega^{a},\omega)\,,\nonumber\\
(\bar{c}^{a},\bar{c})&\leftrightarrow& (\bar{\omega}^{a},\bar{\omega})\,,\nonumber\\
(b^{a},b)&\leftrightarrow& (\bar{b}^{a},\bar{b})\,.  \label{mirrorsymm}
\end{eqnarray}
Moreover, due to the presence of the term $S_{\vartheta}$, expression \eqref{actb} is not left invariant by the BRST transformations which, as in the case of the Landau gauge \cite{Sorella:2010it,Capri:2010pg}, turn out to be softly broken, {\it i.e.}
\begin{equation}
s{\widetilde S} =  sS_{\vartheta}=-i\sqrt{2}\vartheta^{2}\int d^{4}x\,\left[
(\partial_{\mu}c+g\e^{ab}A^{a}_{\mu}c^{b})B_{\mu}
+(\partial_{\mu}\omega+g\e^{ab}B^{a}_{\mu}\omega^{b})A_{\mu}\right]\,,  \label{softb}
\end{equation}
Nevertheless, as we shall see in the following, the presence of the BRST soft breaking does not jeopardize the renormalizability of the model as well as the introduction of local BRST invariant composite operators describing the lightest glueball states. As already argued in  \cite{Sorella:2010it}, the mechanism of the BRST soft breaking could be suitable in order to achieve a good description of the infrared region of a confining gauge theory. On one hand, the soft parameter $\vartheta^2$ induces an infrared modification of the propagators of the elementary gauge fields in such a way that they cannot describe the propagation of physical particles. On the other hand,  a set of suitable composite BRST invariant operators can be introduced in such a way that their correlation functions can be employed to study the physical spectrum of the theory. This framework is precisely what one could expect in a confining theory like pure Yang-Mills theory where gluons are not part of the physical spectrum, which is given by glueballs. \\\\Before addressing the issue of the introduction of the glueball operators, it is worth to elaborate more on the properties of the BRST soft breaking. In particular, in the next section we shall prove that the breaking  \eqref{softterm} can be cast in the form of a linear breaking, {\it i.e.} of a breaking which is purely linear in the quantum fields. This is a relevant property with far reaching consequences. It is known in fact that a linearly broken symmetry can be directly translated into a powerful Ward identity \cite{Piguet:1995er}, which will enable us to give a purely algebraic proof of the renormalizability of the model, as recently done in the case of the Gribov-Zwanziger theory, see  \cite{Capri:2010hb,Capri:2011wp}.

%%%%%%%%%%%%%%%%%%%%%%%%%%%%%%%%%%%%%%%%%%%%%%%%%%%%%%%%%%%
\subsection{Converting the soft breaking into a linear one}
%%%%%%%%%%%%%%%%%%%%%%%%%%%%%%%%%%%%%%%%%%%%%%%%%%%%%%%%%%%
In order to convert the soft breaking, eq.\eqref{softb}, into a linear one, we follow the procedure outlined in   \cite{Capri:2010hb,Capri:2011wp} and introduce a BRST quartet of auxiliary fields
\begin{eqnarray}
&s\lambda=\sigma\,,\qquad s\sigma=0\,,&\nonumber\\
&s\bar{\sigma}=\bar\lambda\,,\qquad s\bar\lambda=0\,.&
\end{eqnarray}
Next, we consider the following action:
\begin{eqnarray}
S_{\sigma}&=&s\int d^{4}x\,\left(\lambda A_{\mu}B_{\mu} -\lambda\bar\sigma +\frac{\zeta_1}{2}\lambda\sigma\right)
 +i\sqrt{2}\vartheta^{2}\int d^{4}x\,\bar\sigma(x)\nonumber\\
 &=&\int d^{4}x\,\biggl(\sigma A_{\mu}B_{\mu}+\frac{\zeta_1}{2}\sigma^{2}
 -\bar{\sigma}(\sigma-i\sqrt{2}\vartheta^{2})\nonumber\\
&& +\lambda\left[\bar{\lambda}+(\partial_{\mu}c+g\e^{ab}A^{a}_{\mu}c^{b})B_{\mu}
+(\partial_{\mu}\omega+g\e^{ab}B^{a}_{\mu}\omega^{b})A_{\mu}\right]\biggl)\,,
\end{eqnarray}
where the term proportional to the dimensionless constant parameter $\zeta_1$ is allowed by power-counting. It is easy to check now that the action $S_{\sigma}$ breaks the BRST symmetry in a linear way, {\it i.e.} the resulting breaking term turns out to be linear in the fields, namely
\begin{equation}
sS_{\sigma}=i\sqrt{2}\vartheta^{2} \int d^{4}x\,\bar\lambda(x)\,.
\end{equation}
It remains now to prove that $S_{\sigma}$ is equivalent to $S_{\vartheta}$. To that end,  let us notice that the auxiliary field $\bar\sigma(x)$ plays the role of  a Lagrange multiplier enforcing the constraint
\begin{equation}
\sigma-i\sqrt{2}\vartheta^{2}=0\,.
\end{equation}
In fact, integrating out the field $\bar{\sigma}$, we have:
\begin{equation}
S_{\sigma}=\int d^{4}x\,\biggl(i\sqrt{2}\vartheta^{2}\,A_{\mu}B_{\mu}-\zeta_1\,\vartheta^{4}
  +\lambda\left[\bar{\lambda}+(\partial_{\mu}c+g\e^{ab}A^{a}_{\mu}c^{b})B_{\mu}
+(\partial_{\mu}\omega+g\e^{ab}B^{a}_{\mu}\omega^{b})A_{\mu}\right]\biggl)\,.
\end{equation}
We can now perform a  change  of variables $(\lambda,\bar\lambda)$:
\begin{eqnarray}
\lambda'&=&\lambda\,,\nonumber\\
{\bar\lambda}'&=&\bar{\lambda}+(\partial_{\mu}c+g\e^{ab}A^{a}_{\mu}c^{b})B_{\mu}
+(\partial_{\mu}\omega+g\e^{ab}B^{a}_{\mu}\omega^{b})A_{\mu}\,.
\end{eqnarray}
with unity Jacobian.  Therefore
\begin{equation}
S_{\sigma}=\int d^{4}x\,\biggl(i\sqrt{2}\vartheta^{2}\,A_{\mu}B_{\mu}-\zeta_1\,\vartheta^{4}
 +\lambda'{\bar\lambda}'\biggr)\,.
\end{equation}
As the last term, $\lambda'{\bar\lambda}'$, is completely decoupled from the rest of theory, we can neglect it. Up to a constant vacuum term, $\zeta_1\,\vartheta^{4}$, we have shown that $S_\sigma$ is equivalent to $S_{\vartheta}$. Thus, the quadratic soft  breaking \eqref{softb} has been  converted into a linear breaking. From now on, we shall  consider the following equivalent action
\begin{equation}
{\widetilde S}_{lin}=S_0[A]+S_0[B]+S_{\sigma}\,. \label{slin}
\end{equation}
Let us end this section by remarking that the equivalence between the two formulations, {\it i.e.} the softly and the linearly broken formulation,  can be promoted at the quantum level. In fact, by following the same steps outlined in the case of the Landau gauge  \cite{Capri:2010hb,Capri:2011wp},  it is not difficult to establish the formal equivalence between the partition functions of the two formulations.
%%%%%%%%%%%%%%%%%%%%%%%%%%%%%%%%%%%%%%%%%%%%%%%%%%%%%%%%%%%%%%%%
\subsection{Introducing dimension two operators}
%%%%%%%%%%%%%%%%%%%%%%%%%%%%%%%%%%%%%%%%%%%%%%%%%%%%%%%%%%%%%%%%
We can pursue now the task of reproducing the infrared behavior of the gluon and ghost
propagators observed in lattice simulations, eqs.\eqref{latt-diag}, \eqref{offd}, \eqref{ghsymm}. The action
\eqref{slin} is not yet in its final form in order to fulfill this task. In fact, we can introduce a whole set of dimension two
operators, $\mathcal{O}^{off}_{A^2}, \mathcal{O}^{diag}_{A^2}, \mathcal{O}_{\bar{c}c}, \mathcal{O}_{\bar{c}\times c}$, which will enable us to reproduce in a nice way the lattice results, while giving rise to an all order renormalizable action. These operators read:
\begin{eqnarray}
\mathcal{O}^{off}_{A^2}&=&\frac{1}{2}(A^{a}_{\mu}A^{a}_{\mu}+B^{a}_{\mu}B^{a}_{\mu})\,,\nonumber \\
\mathcal{O}^{diag}_{A^2}&=&\frac{1}{2}(A_{\mu}A_{\mu}+B_{\mu}B_{\mu})\,, \nonumber \\
\mathcal{O}_{\bar{c}c}&=&\bar{c}^{a}c^{a}+\bar{\omega}^{a}\omega^{a}\,, \nonumber \\
\mathcal{O}_{\bar{c}\times c}&=&g\e^{ab}(\bar{c}^{a}c^{b}+\bar{\omega}^{a}\omega^{b})\,. \label{operators}
\end{eqnarray}
As done in the case of the soft term $S_{\vartheta}$ of eq.\eqref{softterm}, all operators \eqref{operators}
can be introduced in such a way that the resulting breaking of the BRST symmetry is linear. Consider, for example, the
operator $\mathcal{O}^{off}_{A^2}$. Proceeding as in the previous section, we introduce a BRST quartet
\begin{eqnarray}
&s\varphi=\rho\,,\qquad s\rho=0\,,&\nonumber\\
&s\bar{\rho}=\bar\varphi\,,\qquad s\bar\varphi=0\,,&
\end{eqnarray}
and write the following term
\begin{eqnarray}
S_{\rho}&=&s\int d^{4}x\,\left(\varphi\,\mathcal{O}^{off}_{A^{2}}-\varphi\bar\rho +\frac{\zeta_2}{2}\varphi\rho\right)
 +m^{2}\int d^{4}x\,\bar\rho(x)\nonumber\\
 &=&\int d^{4}x\,\biggl(\rho\,\mathcal{O}^{off}_{A^2}+\frac{\zeta_2}{2}\rho^{2}
 -\bar{\rho}(\rho-m^{2})
 +\varphi\left(\bar{\varphi}-s\mathcal{O}^{off}_{A^2}\right)\biggl)\;,  \label{srho}
\end{eqnarray}
where $m$ is a mass parameter and $\zeta_2$ a dimensionless constant allowed by power counting.
Expression $S_{\rho}$ gives rise to a linear breaking of the BRST symmetry, namely
\begin{equation}
s S_{\rho} = m^2 \int d^4x \; \bar\varphi(x) \;. \label{linbsrho}
\end{equation}
Moreover, $S_{\rho}$ is equivalent to the introduction of the term $m^2 \int d^4x\; \mathcal{O}^{off}_{A^2}$. In fact, up to
a constant term, we have
\begin{equation}
S_{\rho} \equiv \int d^{4}x\,\biggl(m^{2}\,\mathcal{O}^{off}_{A^2}+\frac{\zeta_2}{2}m^{4}\biggr)\;, \label{srhoequiv}
\end{equation}
where use has been made of the constraint $\rho =m^2$ and of the change of variable ${\bar \varphi} \rightarrow \bar{\varphi}-s\mathcal{O}^{off}_{A^2}$,
which has unity Jacobian.  The remaining operators can be introduced in the same way. Denoting by $(\psi, h, {\bar h}, {\bar \psi})$ the BRST quartet needed for the introduction of the operator $\mathcal{O}_{\bar{c}c}$, we have
\begin{eqnarray}
&s\psi=h\,,\qquad sh=0\,,&\nonumber\\
&s\bar{h}=\bar\psi\,,\qquad s\bar\psi=0\,,&
\end{eqnarray}
so that the corresponding term accounting for the introduction of $\mathcal{O}_{\bar{c}c}$ is given by
\begin{eqnarray}
S_{h}&=&s\int d^{4}x\,\left(\psi\,\mathcal{O}_{\bar{c}c}-\psi\bar{h} +\frac{\zeta_3}{2}\psi h\right)
 +M^{2}\int d^{4}x\,\bar{h}(x)\nonumber\\
 &=&\int d^{4}x\,\biggl(h\,\mathcal{O}_{\bar{c}c}+\frac{\zeta_3}{2}h^{2}
 -\bar{h}(h-M^{2})
 +\psi\left(\bar{\psi}-s\mathcal{O}_{\bar{c}c}\right)\biggl)\,,\nonumber\\
 &\equiv&\int d^{4}x\,\biggl(M^{2}\,\mathcal{O}_{\bar{c}c}+\frac{\zeta_3}{2}M^{4}\biggr)\,. \label{sh}
\end{eqnarray}
Analogously, for  the operator $\mathcal{O}^{diag}_{A^2}$,  we make use of the quartet
\begin{eqnarray}
&s\chi=\eta\,,\qquad s\eta=0\,,&\nonumber\\
&s\bar{\eta}=\bar\chi\,,\qquad s\bar\chi=0\,,&
\end{eqnarray}
and write the following term
\begin{eqnarray}
S_{\eta}&=&s\int d^{4}x\,\left(\chi\,\mathcal{O}^{diag}_{A^{2}}-\chi\bar\eta +\frac{\zeta_4}{2}\chi\eta\right)
 +\mu^{2}\int d^{4}x\,\bar\eta(x)\nonumber\\
 &=&\int d^{4}x\,\biggl(\eta\,\mathcal{O}^{diag}_{A^2}+\frac{\zeta_4}{2}\eta^{2}
 -\bar{\eta}(\eta-\mu^{2})
 +\chi\left(\bar{\chi}-s\mathcal{O}^{diag}_{A^2}\right)\biggl)\,,\nonumber\\
 &\equiv&\int d^{4}x\,\biggl(\mu^{2}\,\mathcal{O}^{diag}_{A^2}+\frac{\zeta_4}{2}\mu^{4}\biggr)\,. \label{seta}
\end{eqnarray}
Finally, for $\mathcal{O}_{\bar{c}\times c}$, we have
\begin{eqnarray}
&su=\xi\,,\qquad s\xi=0\,,&\nonumber\\
&s\bar{\xi}=\bar{u}\,,\qquad s\bar{u}=0\,,&
\end{eqnarray}
and
\begin{eqnarray}
S_{\xi}&=&s\int d^{4}x\,\left(u\,\mathcal{O}_{\bar{c}\times{c}}-u\bar{h} +\frac{\zeta_5}{2}u\xi\right)
 +v^{2}\int d^{4}x\,\bar{\xi}(x)\nonumber\\
 &=&\int d^{4}x\,\biggl(\xi\,\mathcal{O}_{\bar{c}\times{c}}+\frac{\zeta_5}{2}\xi^{2}
 -\bar{\xi}(\xi-v^{2})
 +u\left(\bar{u}-s\mathcal{O}_{\bar{c}\times{c}}\right)\biggl)\,,\nonumber\\
 &\equiv&\int d^{4}x\,\biggl(v^{2}\,\mathcal{O}_{\bar{c}\times{c}}+\frac{\zeta_5}{2}v^{4}\biggr)\,. \label{sxi}
\end{eqnarray}

%%%%%%%%%%%%%%%%%%%%%%%%%%%%%%%%%%%%%%%%%%%%%%%%%%%
\subsection{The final expression of the classical action for the replica model in the maximal Abelian gauge}
%%%%%%%%%%%%%%%%%%%%%%%%%%%%%%%%%%%%%%%%%%%%%%%%%%%
We have now all ingredients to write down the complete classical action describing the replica model in the maximal Abelian gauge. First, we collect all terms introduced in the previous sections and consider
\begin{eqnarray}
 {\widetilde S}_{lin} +S_\rho+S_{h}+S_\eta+S_{\xi}\,.
\end{eqnarray}
Moreover, in order to discuss the renormalizability of the model, we need to introduce external sources coupled to the composite operators corresponding to the nonlinear BRST transformations \cite{Piguet:1995er}, namely
\begin{equation}
S_{sources}=\int d^{4}x\,\left[\Omega^{a}_{\mu}\,(sA^{a}_{\mu})
+\bar{\Omega}^{a}_{\mu}\,(sB^{a}_{\mu})
+\Omega_{\mu}\,(sA_{\mu})
+\bar{\Omega}_{\mu}\,(sB_{\mu})
+L^{a}\,(sc^{a})+\bar{L}^{a}\,(s\omega^{a})
+L\,(sc)+\bar{L}\,(s\omega)\right]\,.  \label{ssources}
\end{equation}
where $(\Omega^{a}_{\mu}, \bar{\Omega}^{a}_{\mu}, \Omega_{\mu}, \bar{\Omega}_{\mu}, L^{a}, \bar{L}^{a}, L, \bar{L})$ are BRST invariant external fields. It is apparent that the mirror symmetry, eq.\eqref{mirrorsymm},   can be immediately extended to the external sources:
\begin{eqnarray}
(\Omega^{a}_{\mu},\Omega_{\mu})&\leftrightarrow&(\bar\Omega^{a}_{\mu},\bar\Omega_{\mu})\,,\nonumber\\
(L^a,L)&\leftrightarrow&(\bar{L}^{a},\bar{L})\,.
\end{eqnarray}
Finally, it turns out that the following term
\begin{eqnarray}
S_{extra}=\int d^{4}x\,\left(k_1\,\sigma\rho+k_2\,\sigma\eta
+k_3\,\sigma h+k_4\,\rho\eta
+k_5\,\rho h+k_6\,\eta h\right)\,, \label{sextra}
\end{eqnarray}
is required by power-counting. The parameters $(k_1, k_2, k_3,k_4,k_5,k_6)$ will enable us to correctly take into account the possible  mixing arising at the quantum level among the various operators of dimensions two which have been introduced, as it will be discusses in Appendix A. \\\\Therefore, the final form of the classical action describing the replica model in the maximal Abelian gauge is
\begin{equation}
\S= {\widetilde S}_{lin} +S_\rho+S_{h}+S_\eta+S_{\xi} +S_{sources}+S_{extra}\,.\label{sigma}
\end{equation}
It exhibits the important feature of breaking the BRST symmetry in a linear way, {\it i.e.}
\begin{equation}
s \S = \int d^{4}x\,\left(i\sqrt{2}\vartheta^{2}\,\bar\lambda(x)
+m^{2}\,\bar\varphi(x)
+M^{2}\,\bar\psi(x)
+\mu^{2}\,\bar\chi(x)
+v^{2}\,\bar{u}(x)\right)\,.  \label{finalbreak}
\end{equation}
As we shall see in details in Appendix A, eq.\eqref{finalbreak} can be directly translated into a system of helpful Slavnov-Taylor identities.  Moreover, in addition of the Slavnov-Taylor identities, the action  $\S$  exhibits a huge set off additional Ward identities which will ensure the renormalizability to all orders.

\subsection{The gluon and ghost propagators from the replica model and their comparison with the lattice data}

Having identified the final form of the action of the replica model in the maximal Abelian gauge, eq.\eqref{sigma}, let us proceed by evaluating the gluon and ghost propagators and see how they compare with the corresponding lattice expressions, eqs. \eqref{latt-diag}, \eqref{offd}, \eqref{ghsymm}. \\\\ For the transverse off-diagonal gluon component we obtain a Yukawa type propagator
\begin{equation}
\langle A^{a}_{\mu}(k)A^{b}_{\nu}(-k)\rangle=
\langle B^{a}_{\mu}(k)B^{b}_{\nu}(-k)\rangle=
\frac{1}{k^{2}+m^{2}}\left(\d_{\mu\nu}-\frac{k_{\mu}k_{\nu}}{k^{2}}\right)\d^{ab}\,. \label{yk}
\end{equation}
The diagonal gluon propagator turns out to be of the type obtained in \cite{Capri:2008ak}. This kind of propagator also appears in the Refined Gribov-Zwanziger theory of the Landau gauge, see \cite{Dudal:2007cw,Dudal:2008sp,Dudal:2011gd}, {\it i.e.}
\begin{equation}
\langle A_{\mu}(k)A_{\nu}(-k)\rangle=
\langle B_{\mu}(k)B_{\nu}(-k)\rangle=
\frac{k^{2}+\mu^{2}}{(k^{2}+\mu^{2})^{2}+2\vartheta^{4}}\left(\d_{\mu\nu}-\frac{k_{\mu}k_{\nu}}{k^{2}}\right)\,. \label{rgztype}
\end{equation}
For the mixed $A-B$  propagators we get
\begin{equation}
\langle A_{\mu}(k)B_{\nu}(-k)\rangle=
\frac{-2\vartheta^{2}}{(k^{2}+\mu^{2})^{2}+2\vartheta^{4}}\left(\d_{\mu\nu}-\frac{k_{\mu}k_{\nu}}{k^{2}}\right)\,,
\end{equation}
and
\begin{equation}
\langle A^{a}_{\mu}(k)B^{b}_{\nu}(-k)\rangle=
0\,.
\end{equation}
The symmetric off-diagonal ghost propagator is given by
\begin{equation}
\langle \bar{c}^{a}(k)c^{b}(-k)\rangle_{symm}=
\langle \bar{\omega}^{a}(k)\omega^{b}(-k)\rangle_{symm}=
\frac{-k^{2}+M^{2}}{(-k^{2}+M^{2})^{2}+v^{4}}\,\d^{ab}\,, \label{rgzghost}
\end{equation}
while for the anti-symmetric off-diagonal ghost we obtain
\begin{equation}
\langle \bar{c}^{a}(k)c^{b}(-k)\rangle_{anti-symm}=
\langle \bar{\omega}^{a}(k)\omega^{b}(-k)\rangle_{anti-symm}=-
\frac{v^{2}}{(-k^{2}+M^{2})^{2}+v^{4}}\,\e^{ab}\,.
\end{equation}
Looking at expressions \eqref{yk}, \eqref{rgztype}, \eqref{rgzghost} we see that they are precisely of the same qualitative kind of those reported in lattice simulations, and already obtained in \cite{Capri:2008ak}. This  feature shows that the replica model might be very helpful in order to investigate the infrared properties of confining Yang-Mills theories.  Before ending this section, it is worth to spend a few words on the massive parameters $(\vartheta^{2}, m^2, M^2, \mu^2, v^2)$ which appear in the classical action \eqref{sigma}. In the present work, these parameters are regarded as free parameters whose introduction can be supported by the requirement of agreement with the behavior of the gluon and ghost propagators obtained in lattice simulations. Moreover, although being out of the aim of this work, we point out that a possible framework to provide a deeper meaning to those parameters could be achieved by implementing, within the framework of the replica model, the restriction of the domain of integration in the path integral to the so-called Gribov region $\Omega$ of the maximal Abelian gauge\footnote{We remind here that the Gribov region $\Omega$ in the maximal Abelian gauge can be defined as the set of all configurations for which the Faddeev-Popov operator of the maximal Abelian gauge is strictly positive, namely
\begin{equation}
\Omega = \{ A_\mu, A^a_\mu, \;\;\partial_\mu A_\mu=0, \;\;D^{ab}_\mu A^b_\mu=0, \;\;{\cal M}^{ab}>0 \} \;.\label{omegamag}
\end{equation}}. In fact, as it stands, the action of the replica model, eq.\eqref{sigma}, is still plagued by the presence of Gribov copies. This observation follows by realizing that the replica model has been constructed by making use of the maximal Abelian gauge conditions $D^{ab}_\mu A^b_\mu=0$ and ${\bar D}^{ab}_\mu B^b_\mu=0$.  Consequently, the Faddeev-Popov operators  ${\mathcal{M}}^{ab}$  and $\overline{\mathcal{M}}^{ab}$ in eqs.\eqref{fpm},\eqref{fpmo} are nothing but the usual Faddeev-Popov operators of the maximal Abelian gauge. As such, they have zero modes which correspond to Gribov copies. A way to take into account the existence of these copies would be that of  implementing, in the case of the replica model, the restriction in the domain of integration to the Gribov region of the maximal Abelian gauge, as discussed in   \cite{Capri:2005tj,Capri:2006cz,Capri:2008vk,Capri:2010an}, see also \cite{Sorella:2010it}  for a treatment of the Gribov issue in the replica model for the Landau gauge. This would give rise to a set of gap equations which would provide a dynamical origin for the parameters $(\vartheta^{2}, m^2, M^2, \mu^2, v^2)$. Though, for the time being, these parameters will be treated as free parameters.
%%%%%%%%%%%%%%%%%%%%%%%%%%%%%%%%%%%%%%%%%%%%%%%%%%%%
\section{Composite operators and spectral functions}
%%%%%%%%%%%%%%%%%%%%%%%%%%%%%%%%%%%%%%%%%%%%%%%%%%%%
As already pointed out in the introduction, one relevant feature of the replica model is the possibility of introducing BRST invariant composite operators whose two-point correlation functions exhibit the K\"all\'{e}n-Lehmann spectral representation with positive spectral densities. As such, these correlation functions can be given a physical meaning, being suitable for the investigation of the spectrum of the glueballs.\\\\We underline that the introduction of these BRST invariant composite operators is  a nontrivial feature of the replica model, given that these correlation functions have to be evaluated with confining propagators which have complex poles as, for example
\begin{equation}
\frac{k^2 + \mu^2}{(k^2+\mu^2)^2 + 2\vartheta^4} =  \frac{1}{2} \left(   \frac{1}{k^2+\mu^2+i\sqrt{2} \vartheta^2} +   \frac{1}{k^2+\mu^2-i\sqrt{2} \vartheta^2}  \right) \;. \label{ip}
\end{equation}
It is worth mentioning here that this type of propagator also exhibits positivity violation, see  \cite{Dudal:2008sp}. \\\\
Let us also remind that the introduction of BRST invariant composite operators exhibiting good spectral properties is still an open problem of both Gribov-Zwanziger \cite{Zwanziger:1988jt,Zwanziger:1989mf} and Refined Gribov-Zwanziger \cite{Dudal:2007cw,Dudal:2008sp,Dudal:2011gd} frameworks. \\\\The mechanism which enables us to construct composite operators which, despite the use of a confining propagator which has complex poles, give rise to correlation functions with nice spectral properties has been elucidated in \cite{Baulieu:2009ha}  and relies on the concept of $i$-particles. From expression \eqref{ip}, it is apparent that the confining gluon propagator in the right hand side can be seen as describing the propagation of two unphysical modes with complex conjugate masses $(\mu^2 \pm i\sqrt{2} \vartheta^2)$. These unphysical modes have been called $i$-particles in  \cite{Baulieu:2009ha}. Moreover, it turns out that, for a certain type of composite operators, only pairs of $i$-particles with conjugate masses contribute to the two-point correlation functions of these operators. In this case, the complex conjugate poles combine each other in such a way that a cut along the real axis in the complex $k^2$-plane emerges at the end of the computation  \cite{Baulieu:2009ha}, allowing us to write down a nice spectral representation. In the case of the replica model, it turns out that this particular class of composite operators is precisely given by the BRST invariant operators. This is what one would expect in a confining theory: the elementary fields do not correspond to excitations of the physical spectrum, described by a suitable set of BRST invariant composite operators corresponding to bound states of the confined unphysical elementary excitations. \\\\Let us proceed thus by introducing the BRST invariant composite operators with the quantum numbers of the three lightest glueballs. These states are classified according to the values of $J^{PC}$, where $J$ is the angular momentum, $P$ the parity, and $C$ the charge conjugation, see \cite{Mathieu:2008me}
for a recent review on glueballs. The lightest glueball states have $J^{PC}=0^{++}, 0^{-+}, 2^{++}$. Following the same procedure of the replica model of the Landau gauge \cite{Capri:2010pg}, for the composite operators corresponding to $J^{PC}=0^{++}, 0^{-+}, 2^{++}$, we have
\begin{align}
\mathcal{O}_{0^{++}}(x) &=\frac{1}{2}\left(F^A_{\mu\nu}F^A_{\mu\nu}-G^A_{\mu\nu}G^A_{\mu\nu}\right)\\
&=\frac{1}{2} (F^a_{\mu\nu}F^a_{\mu\nu}- G^a_{\mu\nu}G^a_{\mu\nu}+F_{\mu\nu}F_{\mu\nu} -G_{\mu\nu}G_{\mu\nu})
\label{0++}
\end{align}
\begin{align}
\mathcal{O}_{0^{-+}}(x) &=\frac{1}{2}\epsilon_{\mu\nu\rho\sigma}\left(F^A_{\mu\nu}F^A_{\rho\sigma}-G^A_{\mu\nu}G^A_{\rho\sigma}\right) \\
&=\frac{1}{2}\epsilon_{\mu\nu\rho\sigma}(F^a_{\mu\nu}F^a_{\rho\sigma}+F_{\mu\nu}F_{\rho\sigma} - G^a_{\mu\nu}G^a_{\rho\sigma}-G_{\mu\nu}G_{\rho\sigma})
\label{0-+}
\end{align}
\begin{align}
\left[\mathcal{O}_{2^{++}}(x)\right]_{\mu\nu} &=\left(P_{\mu\alpha}P_{\nu\beta}-\frac{1}{3}P_{\mu\nu}P_{\alpha\beta}\right)\left(F^A_{\alpha\sigma}F^A_{\beta\sigma}-G^A_{\alpha\sigma}G^A_{\beta\sigma}\right)  \\  &=\left(P_{\mu\alpha}P_{\nu\beta}-\frac{1}{3}P_{\mu\nu}P_{\alpha\beta}\right)(F^a_{\alpha\sigma}F^a_{\beta\sigma}+F_{\alpha\sigma}F_{\beta\sigma} - G^a_{\alpha\sigma}G^a_{\beta\sigma}-G_{\alpha\sigma}G_{\beta\sigma})
\label{2++}
\end{align}
where $P_{\mu\nu}=\delta_{\mu\nu}\partial^{2}-\partial_{\mu}\partial_{\nu}$ is the transverse projector. As pointed out in \cite{mschaden,Capri:2010pg}, the last operator generates a pure  $2^{++}$, as follows by noticing that expression \eqref{2++} is symmetric, traceless and conserved. \\\\It remains now to evaluate the two-point correlation functions
\begin{align}
\langle {\cal O}_i(k) {\cal O}_i(-k) \rangle  \;, \qquad i=  0^{++}, 0^{-+}, 2^{++} \;, \label{corrf}
\end{align}
and show that they can be cast in the form of a K\"all\'{e}n-Lehmann spectral representation, {\it i.e.}
\begin{align}
\langle {\cal O}_i(k) {\cal O}_i(-k) \rangle = \frac{1}{\pi} \int^{\infty}_{\tau_{0i}} d\tau \; \frac{\rho_i(\tau)}{\tau + k^2}\;. \label{kl}
\end{align}
Making use of $i$-particles, see eq.\eqref{ip}, for the correlation functions  \eqref{corrf} at one-loop order, we get \begin{equation}
\langle\mathcal{O}_{i}(k)\mathcal{O}_{i}(-k)\rangle = 4 \int \frac{d^4p}{(2\pi)^4}
\frac{f_i(p,k-p)}{(p^2 + \mu^2 + i\sqrt{2}\vartheta^2)(p^2 + \mu^2 - i\sqrt{2}\vartheta^2)}   \;, \label{fi}
\end{equation}
where $f_i(p,k-p), i=0^{++}, 2^{++}, 0^{-+}$ are polynomials in the scalar products of the momenta $(k,p)$. There are several ways to cast expression \eqref{fi} in the form of a K\"all\'{e}n-Lehmann representation. As recently discussed in \cite{Dudal:2010wn}, a powerful framework is that of employing Cutkosky's rules, and performing an analytic continuation to complex masses in Euclidean space. Noticing that, at one-loop order, the correlation functions $\langle {\cal O}_i(k) {\cal O}_i(-k) \rangle$ can be decomposed  into diagonal and off-diagonal components, namely
\begin{equation}
\mathcal{O}_i(x) = \mathcal{O}_i^{diag}(x) + \mathcal{O}_i^{off}(x)  \;, \label{dec1}
\end{equation}
\begin{equation}
\langle\mathcal{O}_i(k)\mathcal{O}_i(-k)\rangle = \langle\mathcal{O}_i^{diag}(k)\mathcal{O}_i^{diag}(-k)\rangle
+ \langle\mathcal{O}_i^{off}(k)\mathcal{O}_i^{off}(-k)\rangle \;, \label{dec2}
\end{equation}
it turns out that, after a rather long computation, the corresponding spectral representations are given by the following expressions
\begin{equation}
\langle\mathcal{O}^{off}(k)\mathcal{O}^{off}(-k)\rangle_{0^{++}} = \frac{1}{8\pi^2} \int^{\infty}_{4m^2}\; \frac{d \tau}{\tau +k^2}
\sqrt{1 - \frac{4m^2}{\tau}} (4m^4  + 2\tau^2 + 8m^2(m^2 - \tau)) \;, \label{sp1}
\end{equation}

\begin{equation}
\langle\mathcal{O}^{off}(k)\mathcal{O}^{off}(-k)\rangle_{2^{++}} = \frac{1}{6\pi^2} \int^{\infty}_{4m^2} \; \frac{d \tau}{\tau +k^2}
\sqrt{1 - \frac{4m^2}{\tau}} \;2 \tau^4 (6m^4 -3m^2 \tau +\tau^2)  \;, \label{sp2}
\end{equation}

\begin{equation}
\langle\mathcal{O}^{off}(k)\mathcal{O}^{off}(-k)\rangle_{0^{-+}} = \frac{1}{\pi^2} \int^{\infty}_{4m^2}\; \frac{d \tau}{\tau +k^2}
\sqrt{1 - \frac{4m^2}{\tau}} (\tau^2 - 4m^4 + 4m^2(m^2 - \tau)) \;, \label{sp3}
\end{equation}

\begin{equation}
\langle\mathcal{O}^{diag}(k)\mathcal{O}^{diag}(-k)\rangle_{0^{++}} = \frac{1}{8\pi^2} \int^{\infty}_{2(\mu^2 + \sqrt{\mu^4 + 2\vartheta^4})} \; \frac{d \tau}{\tau +k^2}
\sqrt{1-\frac{8\vartheta^4}{\tau^2} - \frac{4\mu^2}{\tau}} (4\mu^2 + 8\vartheta^4 + 2\tau^2 + 8\mu^2(\mu^2 - \tau)) \;, \label{sp4}
\end{equation}

\begin{eqnarray}
\langle\mathcal{O}^{diag}(k)\mathcal{O}^{diag}(-k)\rangle_{2^{++}} &=& \frac{1}{9\pi^2} \int^{\infty}_{2(\mu^2 + \sqrt{\mu^4 + 2\vartheta^4})} \frac{d \tau}{\tau +k^2}
\sqrt{1-\frac{8\vartheta^4}{\tau^2} - \frac{4\mu^2}{\tau}} \;  \tau^2 \left(-8\tau \vartheta^4 (\mu^2 -4 \tau)\right. \nonumber\\ &&\left. + 3\tau^2 (6\mu^4 -3 \mu^2 \tau +\tau^2) + 32 \vartheta^8    \right)
\;, \label{sp5}
\end{eqnarray}

\begin{equation}
\langle\mathcal{O}^{diag}(k)\mathcal{O}^{diag}(-k)\rangle_{0^{-+}} = \frac{1}{16\pi^2} \int^{\infty}_{2(\mu^2 + \sqrt{\mu^4 + 2\vartheta^4})}  \frac{d \tau}{\tau +k^2}
\sqrt{1-\frac{8\vartheta^4}{\tau^2} - \frac{4\mu^2}{\tau}} (\tau^2 - 4\mu^4 - 8\vartheta^4 + 4\mu^2(\mu^2 - \tau)) \;. \label{sp6}
\end{equation}
It is worth observing that all spectral densities entering expressions \eqref{sp1}--\eqref{sp6} are positive definite within the corresponding range of integration.

%%%%%%%%%%%%%%%%%%%%%%%%%%%%%%%%%%%%%%%%%%%%%%%%%%%%%%%%%%%%%%%%%%%%%%%%%%%%%%%%%%%%%%%%%%%%%%%%%
\section{Qualitative analysis of the spectrum of the lightest glueballs, $J^{PC}= 0^{++},2^{++},0^{-+}$}
%%%%%%%%%%%%%%%%%%%%%%%%%%%%%%%%%%%%%%%%%%%%%%%%%%%%%%%%%%%%%%%%%%%%%%%%%%%%%%%%%%%%%%%%%%%%%%%%%
With the information encoded in the K\"all\'{e}n-Lehmann spectral representation of the correlation functions of the glueball operators, we can provide a qualitative study of the spectrum of the lightest states $J^{PC}= 0^{++},2^{++},0^{-+}$ in the maximal Abelian gauge. As we shall see later on, the hypothesis of Abelian dominance will play a pivotal role. \\\\The first task to study the spectrum of the glueballs is that of establishing a suitable mass formula through the use of the spectral densities evaluated in the previous sections. To that end we shall adopt the set up which has been successfully developed in the case of the Landau gauge \cite{Capri:2010pg}, based on a kind of SVZ-type sum rules \cite{Shifman:1978bx, Shifman:1978by, Shifman:1979if}. For the benefit of the reader, let us give here a short account of the method. As it is customary in the SVZ approach to QCD \cite{Colangelo:2000dp}, we start by considering the two point correlation functions
\begin{equation}
\Pi_i(q^2) = \int d^4x \, e^{iqx} \langle O_i(x) O_i(0) \rangle \;, \label{cr1}
\end{equation}
where $O_i$, $i=0^{++}, 2^{++}, 0^{-+}$, stand for the local composite gauge invariant operators which generate glueball states with quantum numbers $J^{PC} =0^{++}, 2^{++}, 0^{-+}$. \\\\On physical grounds, a truly nonperturbative evaluation of $\Pi_i(q^2)$ would enable us to write an exact K\"all\'{e}n-Lehmann spectral representation
\begin{equation}
\Pi_i(q^2) = \frac{1}{\pi} \int_0^{\infty} d\tau \,\frac{Im \Pi_i(\tau)}{\tau+q^2}  \;, \label{kl}
\end{equation}
which is expected to follow from the unitarity and analyticity properties of the underlying nonperturbative theory\footnote{We remind here that, in some cases,  the spectral representation, eq.\eqref{kl}, might require appropriate subtraction terms in order to ensure convergence. These terms are not written down, as they will be removed once the Borel transformation will be taken \cite{Colangelo:2000dp}.}. We thus proceed by employing a one-resonance parametrization for $Im \Pi_i(\tau)$, namely
\begin{equation}
\frac{Im \Pi_i(\tau)}{\pi} = {\cal R}_i\, \delta(\tau-m^2_i) + \theta(\tau-\tau_0^i) \rho_{i}(\tau) \;, \label{spdec}
\end{equation}
where $m^2_i$ denotes the glueball mass in the $i$-{\it th} channel, $\tau_0^i$ is the threshold for the continuum part of the spectrum, and $\rho_i(\tau)$ the corresponding positive spectral density. Of course, the real values of ${\cal R}_i$, $m^2_i$, $\tau_0^i$ and of the spectral density $\rho_i(\tau)$ are unknown. So far, the best which can be done is computing the correlation functions \eqref{cr1} by trying to encode as much nonperturbative effects as possible, as summarized by the following equation
\begin{equation}
\frac{ {\cal R}_i}{q^2+m^2_i}  + \int_{\tau_0^i}^{\infty} d\tau\, \frac{\rho_{i}(\tau)}{\tau+q^2} =  \Pi_i^{np} \;, \label{sr}
\end{equation}
where $\Pi_i^{np}$ stands for the expression of the correlation function \eqref{cr1} which one has been able to evaluate in practice. Expression \eqref{sr} establishes the so-called sum rules, enabling us to give estimates of the glueball masses in terms of the nonperturbative parameters present in $\Pi_i^{np}$. As done in \cite{Capri:2010pg}, we shall attempt at evaluating the right hand side of eq.\eqref{sr} by employing a suitable trial action $S_{trial}$ which encodes nonperturbative information about the infrared dynamics of Yang-Mills theories. Of course, the choice of $S_{trial}$ is a nontrivial matter which has to account for the following requirements:
\begin{itemize}
\item $i)$ the action  $S_{trial}$ has to display a nonperturbative character, encoded in the presence of a set of nonperturbative parameters $\{ \lambda \}$.  Moreover, in the limit $\lambda=0$, in which these parameters are removed,  the action  $S_{trial}(\lambda)$  has to reduce to the usual perturbative Faddeev-Popov action $S_{FP}$, {\it i.e.}
\begin{equation}
S_{trial}(\lambda) \Bigl|_{\lambda=0}= S_{FP}  \;, \label{fp}
\end{equation}
\item $ii)$ it accounts for gluon confinement. This means that the two point correlation function of the elementary gluon field evaluated with $S_{trial}(\lambda)$  cannot be cast in the form of a spectral representation with positive spectral density, so that it cannot be interpreted as the propagator of a physical particle.
\item $iii)$ $S_{trial}(\lambda)$ has to be renormalizable, meaning that consistent perturbative calculations can be worked out.
\item $iv)$ $S_{trial}(\lambda)$ should  enable us to introduce gauge invariant or, equivalently, BRST invariant local composite operators $\{ O_i\} $ with the quantum numbers $J^{PC} =0^{++}, 2^{++}, 0^{-+}$, whose two point correlation functions
\begin{equation}
\Pi_i^{np}(q^2) \equiv  \Pi_i^{trial}(q^2,\lambda)  = \int d^4x \, e^{iqx} \langle O_i(x) O_i(0) \rangle \;, \label{cr2}
\end{equation}
exhibit the K\"all\'{e}n-Lehmann spectral representation, at least at their lowest order, {\it i.e.}
\begin{equation}
\Pi_i^{np}(q^2) = \Pi_i^{trial}(q^2,\lambda) = \int_{\tau_0^{(i)trial}(\lambda)}^{\infty} d\tau \,\frac{\rho_i^{trial}(\tau)}{\tau+q^2} + {\cal O}(\hbar^2) \;. \label{kl1}
\end{equation}
\end{itemize}
Requirements $i)$--$iv)$ look quite stringent. As such, they might provide a satisfactory set up in order to achieve a useful predictive expression for the correlation functions  $\Pi_i^{trial}(q^2,\lambda)$, eq.\eqref{cr2}. \\\\As trial action we shall adopt the action of the replica model of the maximal Abelian gauge given in the expression \eqref{sigma}, {\it i.e.}
\begin{equation}
S_{trial}(\lambda) = \Sigma \;, \label{trial_act}
\end{equation}
so that the quantity $\Pi_i^{trial}(q^2,\lambda)$ in eq.\eqref{kl1} will be identified with the corresponding expression evaluated with the replica action $\Sigma$, namely
\begin{equation}
\Pi_i^{trial} \equiv \Pi_i^{repl} \;, \qquad  \rho_i^{trial} \equiv \rho_i^{repl}   \;. \label{idt}
\end{equation}
As shown in the previous sections, the action $\Sigma$ fulfills all requirements $i)-iv)$. In particular, the nonperturbative character of $\Sigma$ is encoded in the mass parameters $(\vartheta^2, m^2, M^2, \mu^2, v^2)$ which enable us to reproduce the infrared behavior of the gluon and ghost propagators observed in lattice simulations. Furthermore, taking into account that, according to eqs. \eqref{sp1}-\eqref{sp6}, the spectral functions $\rho_i^{repl}$ decompose at one-loop order as
\begin{equation}
\rho_i^{repl} =  [\rho_i^{repl}]_{diag} + [ \rho_i^{repl}]_{off}  \;, \label{decrho}
\end{equation}
it follows that eq.\eqref{kl1} becomes
\begin{equation}
 \Pi_i^{repl}(q^2) = \int_{[\tau_0^{(i)repl}]_{diag}}^{\infty} d\tau \,\frac{[\rho_i^{repl}(\tau)]_{diag}}{\tau+q^2} +  \int_{[\tau_0^{(i)repl}]_{off}}^{\infty} d\tau \,\frac{[\rho_i^{repl}(\tau)]_{off}}{\tau+q^2}  \;. \label{decPi}
\end{equation}
Before going any further and, as done in the SVZ sum rules \cite{Shifman:1978bx, Shifman:1978by, Shifman:1979if,Colangelo:2000dp}, we need to  provide an estimate for the quantity
\begin{equation}
\int_{\tau_0^i}^{\infty} d\tau\, \frac{\rho_i(\tau)}{\tau+q^2} \;, \label{est}
\end{equation}
appearing in eqs.\eqref{spdec},\eqref{sr}. Requirement $iv)$ above states that the expression for $\Pi_i^{np}$ in eq.\eqref{sr}  reads, at the lowest order:
\begin{equation}
\frac{ {\cal R}_i}{q^2+m^2_i}  + \int_{\tau_0^i}^{\infty} d\tau\, \frac{\rho_i(\tau)}{\tau+q^2} = \int_{[\tau_0^{(i)repl}]_{diag}}^{\infty} d\tau \,\frac{[\rho_i^{repl}(\tau)]_{diag}}{\tau+q^2} +  \int_{[\tau_0^{(i)repl}]_{off}}^{\infty} d\tau \,\frac{[\rho_i^{repl}(\tau)]_{off}}{\tau+q^2}   + {\cal O}(\hbar^2)   \;. \label{fpc}
\end{equation}
Following \cite{Capri:2010pg}, we shall estimate expression \eqref{est} by assuming that, up to the required order, the exact spectral functions $\rho_i(\tau)$ are well approximated by those which have been evaluated explicitly at one-loop order by means of the replica action $\Sigma$, that is
\begin{equation}
\rho_i(\tau) = [\rho_i^{repl}(\tau)]_{diag}\; \theta(\tau - a\; [\tau_0^{(i)repl}]_{diag}) + [\rho_i^{repl}(\tau)]_{off} \;\theta(\tau - b\; [\tau_0^{(i)repl}]_{off}) \;, \label{prm}
\end{equation}
where $a,b \in \mathds{R}^+ $ stand for free parameters accounting for the difference between the value of the  true unknown physical threshold $\tau_0^i$ in eq.\eqref{est}  and the thresholds  $\left( [\tau_0^{(i)repl}]_{diag}, [\tau_0^{(i)repl}]_{off} \right)$ which we have been able to evaluate in practice, eqs.\eqref{sp1}--\eqref{sp6}. Therefore, for the final form of the sum rules which we shall employ, we write
\begin{equation}
\frac{ {\cal R}_i}{q^2+m^2_i} = \int_{ [\tau_0^{(i)repl}]_{diag}}^{ a [\tau_0^{(i)repl}]_{diag}} d\tau \,\frac{[\rho_i^{repl}(\tau)]_{diag}}{\tau+q^2}  +  \int_{ [\tau_0^{(i)repl}]_{off}}^{ b [\tau_0^{(i)repl}]_{off}} d\tau \,\frac{[\rho_i^{repl}(\tau)]_{off}}{\tau+q^2} \;. \label{fsr}
\end{equation}
This equation will be taken as the starting point in order to derive a mass formula for the glueballs and to provide a qualitative analysis of the lightest states $J^{PC} =0^{++}, 2^{++}, 0^{-+}$ in the maximal Abelian gauge.

\subsection{A mass formula for the glueballs}

All ingredients are now at our disposal to work out a mass formula for the glueball states. As in the SVZ sum rules approach \cite{Shifman:1978bx, Shifman:1978by, Shifman:1979if}, we shall make use of the Borel transformation. Let us briefly illustrate how this works. One starts by considering an equation of the following type
\begin{equation}
\frac{\cal F}{k^2+m^2_g} = \int_{\tau_1}^{\tau_2} d\tau \frac{\sigma(\tau)}{\tau + k^2} \;, \label{examp}
\end{equation}
for some suitable ${\cal F}$, $m^2_g$ and $\sigma(\tau)$. Equation \eqref{examp} is precisely of the kind of expression \eqref{fsr}. Let us remind some basic properties of the Borel transformation $\mathcal{B}_\mu$, {\it i.e.}
\begin{align}
\mathcal{B}_\mu \left( \frac{1} {k^2+m^2_g} \right) &= e^{-\frac{m^2_g}{\mu^2_{\cal B}}}
\;,  \notag \\
\mathcal{B}_\mu \left( \frac{k^2} {k^2+m^2_g} \right) & = -m^2_g e^{-\frac{m^2}{\mu^2_{\cal B}}} \;,  \label{br1}
\end{align}
where $\mu_{\cal B}$ is the so called Borel mass, see for example \cite{Colangelo:2000dp}. Applying thus the operation $\mathcal{B}_\mu$  to both sides of eq.\eqref{examp}, we get
\begin{align}
\mathcal{F}e^{-\frac{m^2_g}{\mu^2_{\cal B}}} & =  \int_{\tau_1}^{\tau_2} d\tau \,\sigma(\tau) \;e^{-\frac{%
\tau}{\mu^2_{\cal B}}} \;,  \notag \\
m^2_g \;\mathcal{F}e^{-\frac{m^2_g}{\mu^2_{\cal B}}} & =   \int_{\tau_1}^{\tau_2} d\tau \,\tau \;\sigma(\tau)  \;e^{-%
\frac{\tau}{\mu^2_{\cal B}}}  \label{br3}
\end{align}
so that one derives the mass formula \cite{Capri:2010pg}
\begin{equation}
m^2_g  = \frac{ \int_{\tau_1}^{\tau_2}  d\tau  \,\tau \, \sigma(\tau) \, e^{-\frac{\tau}{\mu^2_{\cal B}}} }{ \int_{\tau_1}^{\tau_2} d\tau \, \sigma(\tau) \, e^{-\frac{\tau}{\mu^2_{\cal B}}} } \;.
\label{massg}
\end{equation}
Let us make use of this formula by focusing, without loss of generality, to the case of the glueball state $0^{++}$. Before applying the Borel transformation, it is useful to perform helpful changes of variables in the integrals of expressions \eqref{sp1} and \eqref{sp4}. For the diagonal part,  eq.\eqref{sp4}, we shall set
\begin{align}
\hat{\tau} &\equiv \tau - 2\mu^2 \;,  \notag \\
\Delta &\equiv 2\sqrt{\mu^4 + 2\vartheta^4}  \;, \label{par1}
\end{align}
so that
\begin{align}
\langle\mathcal{O}^{diag}(k)\mathcal{O}^{diag}(-k)\rangle_{0^{++}} &= \frac{1}{8\pi^2}
\int^{\infty}_{\Delta} d \hat{\tau} \left(\frac{1}{k^2 + \hat{\tau} +2\mu^2}%
\right)\frac{\sqrt{\hat{\tau}^2 - \Delta^2}}{\hat{\tau} + 2\mu^2}\left(2\hat{%
\tau}^2+\Delta^2\right)  \notag \\
&=\frac{1}{8\pi^2} \int^{\infty}_{\Delta} d \hat{\tau} \left(\frac{1}{k^2 +
\hat{\tau} +\sqrt{\Delta^2 - 8\vartheta^4}}\right)\frac{\sqrt{\hat{\tau}^2 -
\Delta^2}}{\hat{\tau} + \sqrt{\Delta^2 - 8\vartheta^4}}\left(2\hat{\tau}%
^2+\Delta^2\right)   \;. \label{diag2}
\end{align}
For the off-diagonal component, eq.\eqref{sp1}, we shall make use of
\begin{align}
\tilde{\tau} &\equiv \tau - 2m^2 \;, \notag \\
\Gamma &\equiv 2m^2  \;.  \label{par2}
\end{align}
Thus
\begin{align}
\langle\mathcal{O}^{off}(k)\mathcal{O}^{off}(-k)\rangle_{0^{++}} = \frac{1}{8\pi^2}
\int^{\infty}_{\Gamma} d \tilde{\tau} \left(\frac{1}{k^2 + \tilde{\tau} +
\Gamma}\right)\frac{\sqrt{\tilde{\tau}^2 - \Gamma^2}}{\tilde{\tau} +\Gamma}%
\left(2\tilde{\tau}^2+\Gamma^2\right)  \;. \label{off2}
\end{align}
Applying now the Borel transformation to the diagonal component, one gets
\begin{align}
\mathcal{B}_\mu \left( \frac{1}{k^2 + \hat{\tau} +\sqrt{\Delta^2 - 8\vartheta^4}%
} \right) &= \mathcal{B}_\mu \left( \frac{1}{k^2 + \hat{\tau} +2\mu^2} \right)
= e^{-\frac{(\hat{\tau} +2\mu^2)}{\mu^2_{\cal B}}} = e^{-pt}e^{-pq}\;,  \notag \\
\mathcal{B}_\mu \left( \frac{k^2}{k^2 + \hat{\tau} + 2\mu^2} \right) &= -\left(%
\hat{\tau} + 2\mu^2\right) e^{-\frac{(\hat{\tau} +2\mu^2)}{\mu^2_{\cal B}}} =
-\Delta\left(t + q\right)e^{-pt}e^{-pq}  \label{br4}
\end{align}
where we have introduced the following dimensionless quantities
\begin{align}
t &= \frac{\hat{\tau}}{\Delta} \;,  \notag \\
p &= \frac{\Delta}{\mu^2_{\cal B}}  \;, \notag \\
q &= \frac{2\mu^2}{\Delta} \;.  \label{par3}
\end{align}
Analogously, for the off-diagonal part:
\begin{align}
\mathcal{B}_\mu \left( \frac{1}{k^2 + \tilde{\tau} + \Gamma} \right) &= e^{-%
\frac{(\tilde{\tau} + \Gamma)}{\mu^2_{\cal B}}} = e^{-rs}e^{-r}\;,  \notag \\
\mathcal{B}_\mu \left( \frac{k^2}{k^2 + \tilde{\tau} + \Gamma} \right) &=
-\left(\tilde{\tau} + \Gamma\right) e^{-\frac{(\tilde{\tau} + \Gamma)}{\mu^2_{\cal B}}}
= -\Gamma\left(s+1\right)e^{-rs}e^{-r}  \label{br5}
\end{align}
where
\begin{align}
s &= \frac{\tilde{\tau}}{\Gamma}  \;, \notag \\
r &= \frac{\Gamma}{\mu^2_{\cal B}}\;.  \label{par4}
\end{align}
Finally, putting all together, we obtain
\begin{align}
m_{0^{++}}^2(a,b,p,q,r) = \Delta \frac{ \int^{a}_{1} d t \sqrt{t^2-1}%
\left(2t^2+1\right) e^{-pt}e^{-pq} + 3\frac{r^4}{p^4} \int^{b}_{1} d s \sqrt{%
s^2-1}\left(2s^2+1\right) e^{-rs}e^{-r} }{ \int^{a}_{1} d t \frac{\sqrt{t^2-1%
}}{t+q}\left(2t^2+1\right) e^{-pt}e^{-pq} + 3\frac{r^3}{p^3} \int^{b}_{1} d
s \frac{\sqrt{s^2-1}}{s+1}\left(2s^2+1\right) e^{-rs}e^{-r} } \;. \label{mform}
\end{align}
This formula follows by performing the following steps: we start from expression  (\ref{fsr}) which is a particular case of expression (\ref{examp}). In particular, the spectral densities appearing in eq.(\ref{fsr}) can be read off from eqs.(\ref{diag2}), (\ref{off2}). Further, the Borel transform in  eq.(\ref{br1}) is applied to the specific case of the spectral spectral densities given in eqs.(\ref{diag2}), (\ref{off2}), resulting in the mass formula \eqref{mform}. Similar expressions can be derived for the masses corresponding to the states $2^{++}, 0^{-+}$.

\subsection{Taking into account the hypothesis of Abelian dominance}

Having been able to work out a mass formula for the glueball states, the next step is that of making a comparison among the three masses $m^2_{0^{++}}, m^2_{2^{++}}, m^2_{0^{-+}}$, and check out if their location is in agreement with the results obtained in lattice simulations \cite{Teper:1998kw}, according to which the $0^{++}$ is the lowest  state, followed by the $2^{++}$, the $0^{-+}$ being the heaviest one, {\it i.e.}  $m^2_{0^{++}} < m^2_{2^{++}} < m^2_{0^{-+}}$. \\\\At this stage, and before discussing the mass hierarchy, we need to take into account the consequences stemming from the hypothesis of Abelian dominance \cite{Ezawa:1982bf,Suzuki:1989gp,Hioki:1991ai}, which is believed to be a key feature of the maximal Abelian gauge. Let us remind here that, according to the Abelian dominance, the off-diagonal degrees of freedom are expected to decouple at low energies, due to the dynamical generation of a large off-diagonal mass, a fact which has already received a certain amount of evidence, from numerical lattice simulations \cite{Amemiya:1998jz,Bornyakov:2003ee,Mendes:2006kc} as well as from theoretical calculations \cite{Schaden:1999ew,Kondo:2001tm,Dudal:2004rx}. Even if the issue of the Abelian dominance is still an open subject, let us proceed by assuming that it holds and that, accordingly, the off-diagonal components acquire a large dynamical mass which decouple them at low energies. As we shall see, the hypothesis of the Abelian dominance will have  great impact on the glueball mass formula \eqref{mform}. Looking in fact at expression \eqref{mform}, one immediately recognizes that the contribution coming from the off-diagonal sector is encoded in the two integrals
\begin{equation}
\frac{r^4}{p^4} \int_{1}^{b} ds \; \sqrt{s^2-1}\; e^{-rs} \; e^{-r}  \;, \qquad
\frac{r^3}{p^3} \int_{1}^{b} ds \; \frac{\sqrt{s^2-1}}{s+1} \; (2s^2+1)\; e^{-rs} \; e^{-r}  \;,  \label{2int}
\end{equation}
where
\begin{equation}
r = \frac{\Gamma}{\mu^2_{\cal B}} = 2 \frac{m^2}{\mu^2_{\cal B}}  \;, \label{rdef}
\end{equation}
and $m$ stands for the mass of the off-diagonal components of the gauge field. If we assume now that, according to the Abelian dominance hypothesis, $m$ is larger enough than the remaining mass parameters $(\mu, \vartheta)$ which enter the expression of the diagonal gluon propagator, eq.\eqref{rgztype}, it follows that the off-diagonal contribution to expression \eqref{mform} turns out to be very much suppressed, due to the exponential factor $e^{-r(s+1)}$ appearing in eq.\eqref{2int}. Neglecting thus the off-diagonal contribution, the mass formula for the state $0^{++}$ becomes
\begin{equation}
m^{2}_{0^{++}}  {\rightarrow}_{Abelian{\;} dom.} = \Delta \;\frac{ \int_{1}^{a} dt \sqrt{t^2-1}( t^2 + \frac{1}{2} ) \; e^{-pt}} {\int_{1}^{a} dt \frac{\sqrt{t^2-1}}{t+q} (t^2 + \frac{1}{2}) e^{-pt} } \;. \label{abd0++}
\end{equation}
Similarly, for the other states $2^{++}$ and $0^{++}$, we obtain
\begin{eqnarray}
m^2_{2^{++}}(a,p) & =& \Delta\;  \frac{ \int_{1}^{a} dt \;\sqrt{t^2-1}\, (q+t)^2 (5 q^2 (2t^2+1) + 15 qt (t^2+1) + 6t^4 +8t +1) \, e^{-pt}       }
{ \int_{1}^{a} dt \;\frac{\sqrt{t^2-1}}{t+q} \,  (q+t)^2 (5 q^2 (2t^2+1) + 15 qt (t^2+1) + 6t^4 +8t +1) \, e^{-pt}  } \;, \label{m2++a} \\[5mm]
m^2_{0^{-+}}(a,p) & =& \Delta \frac{ \int_{1}^{a} dt \;\left(t^2-1\right)^{3/2} \, e^{-pt}        } { \int_{1}^{a} dt \frac{\left(t^2-1\right)^{3/2}} {t+q}\, e^{-pt}   } \;. \label{m0-+a}
\end{eqnarray}
As done in \cite{Capri:2010pg}, we present here a qualitative analysis of the ratio of the glueball masses $m^{2}_{0++}$, $m^{2}_{2++}$, $m^{2}_{0-+}$.  From expressions \eqref{abd0++}, \eqref{m2++a} and \eqref{m0-+a} we see that, under the hypothesis of Abelian dominance, the glueball masses are functions of the threshold parameter $a$ and of the Borel mass $\mu_{\cal B}$, encoded in the parameter $p$. According to the SVZ framework, we vary these parameters and we look at the location of the glueball masses. The output of our results are shown
in Fig.1 and Fig.2. In particular, from Fig.1, one can see that, when the parameter $a$ belongs to the interval
$1 < a < 1.8$, the masses of the three lightest states are in qualitative agreement with the available lattice data \cite{Teper:1998kw},
{\it i.e.} $m^{2}_{0++} < m^{2}_{2++} < m^{2}_{0-+}$, a feature which holds for all values of $p$, as shown in Fig.2. As already mentioned, the plots of Fig.1 and Fig. 2 are precisely of the  same kind of those previously obtained in the Landau gauge \cite{Capri:2010pg}, a fact which we interpret as evidence of the gauge independence of the physical spectrum of the theory. Moreover, the existence of the interval $1 < a < 1.8$ is regarded as an encouraging consistent confirmation towards a more quantitative analysis of the glueballs spectrum in the maximal Abelian gauge \cite{prep}, along the lines presented in \cite{Dudal:2010cd}.

\begin{figure}[H]
\label{fig:0pp2pp0mp}
  \centering
  {\includegraphics[scale=1]{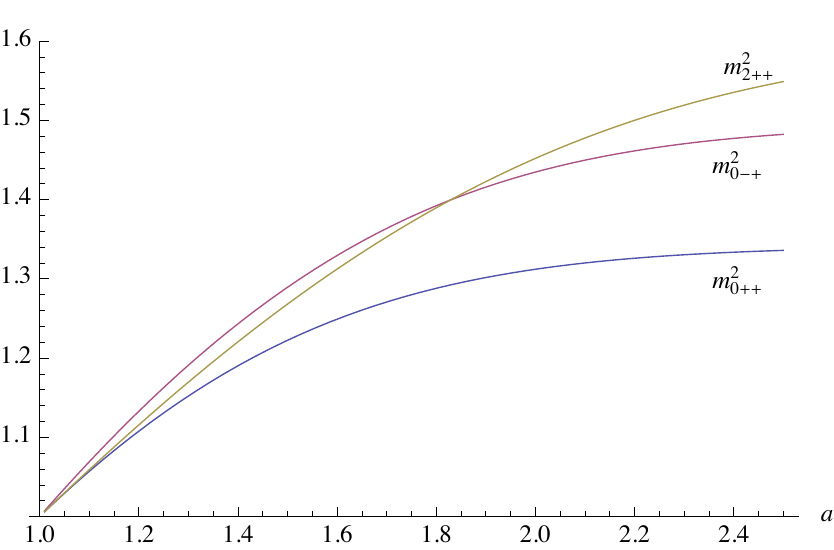}}
  \caption{Glueball masses as functions of the threshold parameter $a$, for $p=5$.}
\end{figure}

\begin{figure}[H]
\label{fig:0pp2pp0mp-p}
 \centering
 {\includegraphics[scale=1]{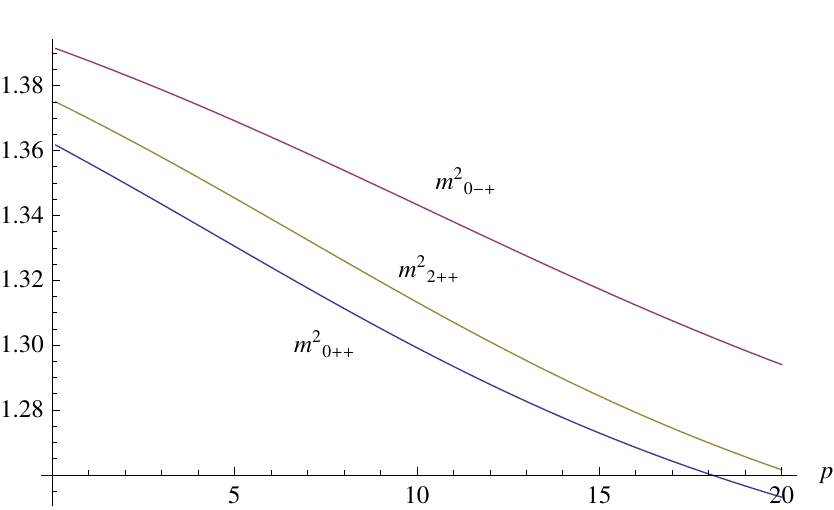}}
 \caption{Glueball masses as functions of $p$, for $a=1.3$}
\end{figure}

\noindent Let us end this section with the following remarks:
\begin{itemize}
\item it turns out that the mass hierarchy for the states $0^{++}, 2^{++}, 0^{-+}$, in the maximal Abelian gauge is the same as that of the Landau gauge, namely $m^2_{0^{++}} < m^2_{2^{++}} < m^2_{0^{-+}}$, and is in agreement with the lattice data \cite{Teper:1998kw}. We take the present results as evidence of the gauge independence of the physical spectrum of the theory,

\item it worth to mention that the ordering which we have obtained for the low lying glueball masses has also been established in other approaches, such as AdS/QCD and SVZ sum rules. We remind the reader to the reference \cite{Mathieu:2008me} for  a general review on  the glueball spectrum.

\item the present result can be traced back to two relevant features of the maximal Abelian gauge: $i)$ the assumption of the Abelian dominance \cite{Ezawa:1982bf,Suzuki:1989gp,Hioki:1991ai},  $ii)$ the nice fact that the diagonal gluon propagator, eq.\eqref{rgztype}, is of the same kind of that  found in the Landau gauge,  in both  Refined Gribov-Zwanziger model \cite{Dudal:2007cw,Dudal:2008sp,Dudal:2011gd}  and  Landau replica model \cite{Sorella:2010it}.

\end{itemize}

%%%%%%%%%%%%%%%%%%%%
\section{Conclusion}
%%%%%%%%%%%%%%%%%%%%
In this work we have presented a first qualitative study of the glueball spectrum in the maximal Abelian gauge. In particular, the three lightest states, $J^{PC}=0^{++}, 2^{++}, 0^{-+}$, have been investigated. The output of our analysis is that the resulting mass hierarchy is in agreement with both lattice data  \cite{Teper:1998kw} and  previous analytic results obtained in the Landau gauge \cite{Capri:2010pg,Dudal:2010cd}. As already mentioned, we interpret our result as evidence of the gauge independence of the physical spectrum of the theory. \\\\Concerning the maximal Abelian gauge, we have implemented it within the framework of the replica model, already successfully introduced in the Landau gauge \cite{Sorella:2010it}. As we have seen, the action of the replica model which we have been able to construct in the maximal Abelian gauge enables us to nicely reproduce the infrared behavior of the gluon and ghost propagators observed in lattice simulations \cite{Mendes:2006kc}. In addition, the model allows for the introduction of BRST invariant local operators for glueball states whose correlation functions display a spectral representation with positive spectral densities. \\\\We have had also the opportunity to repeatedly state the relevance of the Abelian dominance hypothesis \cite{Ezawa:1982bf,Suzuki:1989gp,Hioki:1991ai}, which is believed to be a key ingredient of the maximal Abelian gauge. We hope that the present results might stimulate further investigations on the infrared behavior of the gluon and ghost propagators through lattice simulations as well as more quantitative results on the Abelian dominance in the maximal Abelian gauge.

%%%%%%%%%%%%%%%%%%%%%%%%%%
\section*{Acknowledgments}
%%%%%%%%%%%%%%%%%%%%%%%%%%%
The Conselho Nacional de Desenvolvimento Cient\'{\i}fico e
Tecnol\'{o}gico (CNPq-Brazil), the Faperj, Funda{\c{c}}{\~{a}}o de
Amparo {\`{a}} Pesquisa do Estado do Rio de Janeiro, the Latin
American Center for Physics (CLAF), the SR2-UERJ,  the
Coordena{\c{c}}{\~{a}}o de Aperfei{\c{c}}oamento de Pessoal de
N{\'{\i}}vel Superior (CAPES)  are gratefully acknowledged.

\appendix
%%%%%%%%%%%%%%%%%%%%%%%%%
\section{Renormalization of the action of the replica model in the maximal Abelian gauge}
%%%%%%%%%%%%%%%%%%%%%%%%%
%%%%%%%%%%%%%%%%%%%%%%%%%%%%%%%%%%%%%%
\subsection{The starting classical action}
%%%%%%%%%%%%%%%%%%%%%%%%%%%%%%%%%%%%%%
This Appendix is devoted to give a detailed proof of the renormalizability of the action of the replica model in the maximal Abelian gauge. Let us begin by reminding the expression of the classical action $\Sigma$ we start with, namely
\begin{equation}
\S= {\widetilde S}_{lin} +S_\rho+S_{h}+S_\eta+S_{\xi} +S_{sources}+S_{extra}\,.\label{sigmaapp}
\end{equation}
where ${\widetilde S}_{lin}, S_\rho, S_{h}, S_\eta, S_{\xi},  S_{sources}, S_{extra}$ are given, respectively, in eqs.\eqref{slin},\eqref{srho},\eqref{sh},\eqref{seta},\eqref{sxi},\eqref{ssources},\eqref{sextra}. With the exception of a linear term, irrelevant at the quantum level, the action $\Sigma$ displays the following discrete symmetry
\begin{eqnarray}
\mathcal{Y}^{1}&\to&-\mathcal{Y}^{1}\,,\nonumber\\
\mathcal{Y}^{2}&\to&\mathcal{Y}^{2}\,,\nonumber\\
\mathcal{Y}^{diag}&\to&-\mathcal{Y}^{diag}\,,\nonumber\\
\mathcal{Y}&\to&\mathcal{Y}\,,\nonumber\\
p&\to&p\,,
\end{eqnarray}
where $ \mathcal{Y}^{a}, a=1,2$, $\mathcal{Y}^{diag}$,  $\mathcal{Y}$ and $p$ stand for
\begin{eqnarray}
\mathcal{Y}^{a}&\equiv&\{A^{a}_{\mu},B^{a}_{\mu},b^{a},\bar{b}^{a},c^{a},\omega^{a},\bar{c}^{a},\bar\omega^{a},\Omega^{a}_{\mu},\bar\Omega^{a}_{\mu},
L^{a},\bar{L}^{a}\}\,,\nonumber\\
\mathcal{Y}^{diag}&\equiv&\{A_{\mu},B_{\mu},b,\bar{b},c,\omega,\bar{c},\bar\omega,\Omega_{\mu},\bar\Omega_{\mu},L,\bar{L}, u,\bar{u}, \xi,\bar{\xi}\}\,,\nonumber\\
\mathcal{Y}&\equiv&\{\sigma,\bar\sigma,\lambda,\bar\lambda,\rho,\bar\rho,\varphi,\bar\varphi,h,\bar{h},\psi,\bar\psi,\eta,\bar\eta,\chi,\bar\chi\}\,,\nonumber\\
p&\equiv&\{m^{2},M^{2},\mu^{2},\vartheta^{2},v^{2},g,\alpha,\zeta_{1}, \dots,\zeta_{5},k_{1},\dots,k_{6}\}\,.
\end{eqnarray}
This symmetry forbids, for example, the occurrence of terms like $\int d^{4}x\,\sigma(x)\xi(x)$ which are, in principle,  permitted by the power-counting.
%%%%%%%%%%%%%%%%%%%%%%%%%%%%
\subsection{Ward identities}
%%%%%%%%%%%%%%%%%%%%%%%%%%%%
The action $\Sigma$ obeys a large set of Ward identities, which we enlist below:
\begin{itemize}
\item{the linearly broken Slavnov-Taylor identity:
\begin{eqnarray}
\mathcal{S}(\S)&\equiv&\int d^{4}x\,\biggl(\frac{\d\S}{\d A^{a}_{\mu}}\frac{\d\S}{\d \Omega^{a}_{\mu}}
+\frac{\d\S}{\d B^{a}_{\mu}}\frac{\d\S}{\d \bar\Omega^{a}_{\mu}}
+\frac{\d\S}{\d A_{\mu}}\frac{\d\S}{\d \Omega_{\mu}}
+\frac{\d\S}{\d B_{\mu}}\frac{\d\S}{\d \bar\Omega_{\mu}}
+\frac{\d\S}{\d c^{a}}\frac{\d\S}{\d L^{a}}
+\frac{\d\S}{\d \omega^{a}}\frac{\d\S}{\d \bar{L}^{a}}\nonumber\\
&&+\frac{\d\S}{\d c}\frac{\d\S}{\d L}
+\frac{\d\S}{\d \omega}\frac{\d\S}{\d \bar{L}}
+ib^{a}\frac{\d\S}{\d \bar{c}^{a}}
+i\bar{b}^{a}\frac{\d\S}{\d \bar{\omega}^{a}}
+ib\frac{\d\S}{\d \bar{c}}
+i\bar{b}\frac{\d\S}{\d \bar{\omega}}
+\s\frac{\d\S}{\d\lambda}
+\bar{\lambda}\frac{\d\S}{\d\bar{\sigma}}\nonumber\\
&&+\rho\frac{\d\S}{\d\varphi}
+\bar{\varphi}\frac{\d\S}{\d\bar{\rho}}
+h\frac{\d\S}{\d\psi}
+\bar{\psi}\frac{\d\S}{\d\bar{h}}
+\eta\frac{\d\S}{\d\chi}
+\bar{\chi}\frac{\d\S}{\d\bar{\eta}}
+\xi\frac{\d\S}{\d u}
+\bar{u}\frac{\d\S}{\d\bar{\xi}}\biggr)\nonumber\\
&=&\int d^{4}x\,\left(i\sqrt{2}\vartheta^{2}\,\bar\lambda(x)
+m^{2}\,\bar\varphi(x)
+M^{2}\,\bar\psi(x)
+\mu^{2}\,\bar\chi(x)
+v^{2}\,\bar{u}(x)\right)\,.
\end{eqnarray}}
\item{the equations of motion of the diagonal Lagrange multipliers and of the antighost equations:
\begin{equation}
\frac{\d\S}{\d{b}}=i\p_{\mu}A_{\mu}\,,\qquad
\frac{\d\S}{\d\bar{c}}+\p_{\mu}\frac{\d\S}{\d\Omega_{\mu}}=0\,; \label{lm1}
\end{equation}
\begin{equation}
\frac{\d\S}{\d{\bar{b}}}=i\p_{\mu}B_{\mu}\,,\qquad
\frac{\d\S}{\d\bar{\omega}}+\p_{\mu}\frac{\d\S}{\d\bar\Omega_{\mu}}=0\,. \label{lm2}
\end{equation}}
\item{the equations of motion of the auxiliary fields:
\begin{equation}
\frac{\d\S}{\d{\bar{\sigma}}}=-(\s-i\sqrt{2}\vartheta^{2})\,,\qquad
\frac{\d\S}{\d{\bar{\lambda}}}=-\lambda\,; \label{la1}
\end{equation}
\begin{equation}
\frac{\d\S}{\d{\bar{\rho}}}=-(\rho-m^{2})\,,\qquad
\frac{\d\S}{\d{\bar{\varphi}}}=-\varphi\,; \label{la2}
\end{equation}
\begin{equation}
\frac{\d\S}{\d{\bar{h}}}=-(h-M^{2})\,,\qquad
\frac{\d\S}{\d{\bar{\psi}}}=-\psi\,; \label{la3}
\end{equation}
\begin{equation}
\frac{\d\S}{\d{\bar{\eta}}}=-(\eta-\mu^{2})\,,\qquad
\frac{\d\S}{\d{\bar{\chi}}}=-\chi\,; \label{la4}
\end{equation}
\begin{equation}
\frac{\d\S}{\d{\bar{\xi}}}=-(\xi-v^{2})\,,\qquad
\frac{\d\S}{\d{\bar{u}}}=-u\,. \label{la5}
\end{equation}}
\item{the linearly broken parametric equations:
\begin{eqnarray}
\frac{\p\S}{\p{\vartheta^{2}}}&=&i\sqrt{2}\int d^{4}x\,\bar\s(x)\,;\\
\frac{\p\S}{\p{m^{2}}}&=&\int d^{4}x\,\bar\rho(x)\,;\\
\frac{\p\S}{\p{M^{2}}}&=&\int d^{4}x\,\bar{h}(x)\,;\\
\frac{\p\S}{\p{\mu^{2}}}&=&\int d^{4}x\,\bar\eta(x)\,;\\
\frac{\p\S}{\p{v^{2}}}&=&\int d^{4}x\,\bar\xi(x)\,.  \label{peqs}
\end{eqnarray}
}
\item{the linearly broken integrated Ward identity:
\begin{eqnarray}
\Upsilon(\S)&\equiv&\int d^{4}x\,\biggl(\frac{\d\S}{\d\varphi}
-\alpha\frac{\d\S}{\d\psi}
-ic^{a}\frac{\d\S}{\d b^{a}}
-i\omega^{a}\frac{\d\S}{\d\bar{b}^{a}}
+2\frac{\d\S}{\d\bar{u}}\frac{\d\S}{\d{L}}
+2\frac{\d\S}{\d\bar{u}}\frac{\d\S}{\d{\bar{L}}}\biggr)\nonumber\\
&=&\int d^4x\,\left(\bar\varphi -\alpha\bar\psi\right)\,.
\end{eqnarray}
}
\item{the diagonal local ghost equations:
\begin{eqnarray}
G_{c}(\S)&\equiv&\frac{\d\S}{\d{c}}
-ig\e^{ab}\bar{c}^{a}\frac{\d\S}{\d{b}^{b}}
+\p_\mu\biggl(A_{\mu}\frac{\d\S}{\d\bar\chi}\biggr)\nonumber\\
&=&-\p_{\mu}(\p_{\mu}\bar{c}+\Omega_\mu)+g\e^{ab}(\Omega^{a}_{\mu}A^{b}_{\mu}-L^{a}c^{b})\,;
\end{eqnarray}
\begin{eqnarray}
G_{\omega}(\S)&\equiv&\frac{\d\S}{\d{\omega}}
-ig\e^{ab}\bar{\omega}^{a}\frac{\d\S}{\d\bar{b}^{b}}
+\p_\mu\biggl(B_{\mu}\frac{\d\S}{\d\bar\chi}\biggr)\nonumber\\
&=&-\p_{\mu}(\p_{\mu}\bar{\omega}+\bar\Omega_\mu)+g\e^{ab}(\bar\Omega^{a}_{\mu}B^{b}_{\mu}-\bar{L}^{a}\omega^{b})\,.
\end{eqnarray}
}
\item{the $SL(2,\Re)$ Ward identities:
\begin{eqnarray}
\mathcal{D}(\S)&\equiv&\int d^{4}x\,\biggl(
c^{a}\frac{\d\S}{\d\bar{c}^{a}}
-i\frac{\d\S}{\d{L}^{a}}\frac{\d\S}{\d{b}^{a}}
+2\frac{\d\S}{\d{L}}\frac{\d\S}{\d\bar\xi}\biggr)=0\,;\\
\bar{\mathcal{D}}(\S)&\equiv&\int d^{4}x\,\biggl(
\omega^{a}\frac{\d\S}{\d\bar{\omega}^{a}}
-i\frac{\d\S}{\d\bar{L}^{a}}\frac{\d\S}{\d\bar{b}^{a}}
+2\frac{\d\S}{\d\bar{L}}\frac{\d\S}{\d\bar\xi}\biggr)=0\,.
\end{eqnarray}
}
\item{the linearly broken $U(1)$ residual identities, typical of the maximal Abelian gauge:
\begin{eqnarray}
\mathcal{W}(\S)&\equiv&\p_{\mu}\frac{\d\S}{\d{A}_{\mu}}
+\p_{\mu}\biggl(A_{\mu}\frac{\d\S}{\d\bar\eta}\biggr)
+g\e^{ab}\,\sum_{y\in\mathcal{A}}y^{a}\frac{\d\S}{\d y^{b}}\nonumber\\
&=&-i\p^{2}b+\mu^{2}\p_{\mu}A_{\mu}\,;
\end{eqnarray}
\begin{eqnarray}
\bar{\mathcal{W}}(\S)&\equiv&\p_{\mu}\frac{\d\S}{\d{B}_{\mu}}
+\p_{\mu}\biggl(B_{\mu}\frac{\d\S}{\d\bar\eta}\biggr)
+g\e^{ab}\,\sum_{\bar{y}\in\mathcal{B}}\bar{y}^{a}\frac{\d\S}{\d \bar{y}^{b}}\nonumber\\
&=&-i\p^{2}\bar{b}+\mu^{2}\p_{\mu}B_{\mu}\,;
\end{eqnarray}
with
\begin{eqnarray}
\mathcal{A}&\equiv&\{A^{a}_{\mu}, b^{a}, \bar{c}^{a}, c^{a}, \Omega^{a}_{\mu}, L^{a}\}\,,\nonumber\\
\mathcal{B}&\equiv&\{B^{a}_{\mu},\bar{b}^{a}, \bar{\omega}^{a}, \omega^{a}, \bar\Omega^{a}_{\mu}, \bar{L}^{a}\}\,.
\end{eqnarray}
}
\end{itemize}
%%%%%%%%%%%%%%%%%%%%%%%%%%%%%%%%%%%%%%%%%
\subsection{Characterization of the  most general counterterm}
%%%%%%%%%%%%%%%%%%%%%%%%%%%%%%%%%%%%%%%%%
In order to characterize the most general local invariant counterterm compatible with the whole set of Ward identities we follow the procedure of the Algebraic Renormalization \cite{Piguet:1995er}. We perturb thus the action $\S$ by adding an arbitrary local integrated polynomial $\S^{count}$ in the fields and external sources, of dimension bounded by four, and we require that the perturbed action, $\S+\e\S^{count}$, obeys, to the first order in the expansion parameter $\e$, the same set of Ward identities fulfilled by $\S$. This gives rise to the following constraints for $\S^{count}$:
\begin{equation}
\mathcal{S}_{\Sigma}\,\S^{count} =0\,;
\end{equation}
\begin{equation}
\left(\frac{\d}{\d\bar{c}}+\p_{\mu}\frac{\d}{\d\Omega_{\mu}}\right)\S^{count}=0\,,\qquad
\left(\frac{\d}{\d\bar{\omega}}+\p_{\mu}\frac{\d}{\d\bar\Omega_{\mu}}\right)\S^{count}=0\,, \label{ct1}
\end{equation}
\begin{equation}
\Upsilon_{\S}\,\S^{count}=0\,;  \label{ct2}
\end{equation}
\begin{equation}
G_c\,\S^{count}=0\,,\qquad
G_\omega\,\S^{count}=0\,;  \label{ct3}
\end{equation}
\begin{equation}
\mathcal{D}_{\S}\,\S^{count}=0\,,\qquad
\bar{\mathcal{D}}_{\S}\,\S^{count}=0\,;  \label{ct4}
\end{equation}
\begin{equation}
\mathcal{W}\,\S^{count}=0\,,\qquad
\bar{\mathcal{W}}\,\S^{count}=0\,;  \label{ct5}
\end{equation}
where $\mathcal{S}_{\S}$ stands for the linearized nilpotent Slavnov-Taylor operator
\begin{eqnarray}
\mathcal{S}_{\S}&=&\int d^{4}x\,\biggl(\frac{\d\S}{\d A^{a}_{\mu}}\frac{\d}{\d \Omega^{a}_{\mu}}
+\frac{\d\S}{\d \Omega^{a}_{\mu}}\frac{\d}{\d A^{a}_{\mu}}
+\frac{\d\S}{\d B^{a}_{\mu}}\frac{\d}{\d \bar\Omega^{a}_{\mu}}
+\frac{\d\S}{\d \bar\Omega^{a}_{\mu}}\frac{\d}{\d B^{a}_{\mu}}
+\frac{\d\S}{\d A_{\mu}}\frac{\d}{\d \Omega_{\mu}}
+\frac{\d\S}{\d \Omega_{\mu}}\frac{\d}{\d A_{\mu}}
+\frac{\d\S}{\d B_{\mu}}\frac{\d}{\d \bar\Omega_{\mu}}
+\frac{\d\S}{\d \bar\Omega_{\mu}}\frac{\d}{\d B_{\mu}}\nonumber\\
&&
+\frac{\d\S}{\d c^{a}}\frac{\d}{\d L^{a}}
+\frac{\d\S}{\d L^{a}}\frac{\d}{\d c^{a}}
+\frac{\d\S}{\d \omega^{a}}\frac{\d}{\d \bar{L}^{a}}
+\frac{\d\S}{\d \bar{L}^{a}}\frac{\d}{\d \omega^{a}}
+\frac{\d\S}{\d c}\frac{\d}{\d L}
+\frac{\d\S}{\d L}\frac{\d}{\d c}
+\frac{\d\S}{\d \omega}\frac{\d}{\d \bar{L}}
+\frac{\d\S}{\d \bar{L}}\frac{\d}{\d \omega}
+ib^{a}\frac{\d}{\d \bar{c}^{a}}
+i\bar{b}^{a}\frac{\d}{\d \bar{\omega}^{a}}
\nonumber\\
&&
+ib\frac{\d}{\d \bar{c}}
+i\bar{b}\frac{\d}{\d \bar{\omega}}
+\s\frac{\d}{\d\lambda}
+\bar{\lambda}\frac{\d}{\d\bar{\sigma}}
+\rho\frac{\d}{\d\varphi}
+\bar{\varphi}\frac{\d}{\d\bar{\rho}}
+h\frac{\d}{\d\psi}
+\bar{\psi}\frac{\d}{\d\bar{h}}
+\eta\frac{\d}{\d\chi}
+\bar{\chi}\frac{\d}{\d\bar{\eta}}
+\xi\frac{\d}{\d u}
+\bar{u}\frac{\d}{\d\bar{\xi}}\biggr)\,,\label{linearized}
\end{eqnarray}
\begin{equation}
\mathcal{S}_{\S} \mathcal{S}_{\S} =0 \;, \label{linst}
\end{equation}
and $\Upsilon_{\S}, {\mathcal{D}}_{\S}, \bar{\mathcal{D}}_{\S}$ are given by
\begin{eqnarray}
\Upsilon_{\S}&=&\int d^{4}x\,\biggl(\frac{\d}{\d\varphi}
-\alpha\frac{\d}{\d\psi}
-ic^{a}\frac{\d}{\d b^{a}}
-i\omega^{a}\frac{\d}{\d\bar{b}^{a}}
+2\frac{\d\S}{\d\bar{u}}\frac{\d}{\d{L}}
+2\frac{\d\S}{\d{L}}\frac{\d}{\d\bar{u}}
+2\frac{\d\S}{\d\bar{u}}\frac{\d}{\d{\bar{L}}}
+2\frac{\d\S}{\d{\bar{L}}}\frac{\d}{\d\bar{u}}\biggr)\,,
\end{eqnarray}
\begin{eqnarray}
\mathcal{D}_{\S}&=&\int d^{4}x\,\biggl(
c^{a}\frac{\d}{\d\bar{c}^{a}}
-i\frac{\d\S}{\d{L}^{a}}\frac{\d}{\d{b}^{a}}
-i\frac{\d\S}{\d{b}^{a}}\frac{\d}{\d{L}^{a}}
+2\frac{\d\S}{\d{L}}\frac{\d}{\d\bar\xi}
+2\frac{\d\S}{\d\bar\xi}\frac{\d}{\d{L}}\biggr)\,,\\
\bar{\mathcal{D}}_{\S}&=&\int d^{4}x\,\biggl(
\omega^{a}\frac{\d}{\d\bar{\omega}^{a}}
-i\frac{\d\S}{\d\bar{L}^{a}}\frac{\d}{\d\bar{b}^{a}}
-i\frac{\d\S}{\d\bar{b}^{a}}\frac{\d}{\d\bar{L}^{a}}
+2\frac{\d\S}{\d\bar{L}}\frac{\d}{\d\bar\xi}
+2\frac{\d\S}{\d\bar\xi}\frac{\d}{\d\bar{L}}\biggr)\,.
\end{eqnarray}
From the equations of motion of the Lagrange multipliers, eqs.\eqref{lm1},\eqref{lm2}, and from those of the auxiliary fields, eqs.\eqref{la1}-\eqref{la5}, it follows immediately that the counterterm $\S^{count}$ must be independent from the set of fields
\begin{equation}
\{b,\bar{b}, \bar\sigma,\bar\lambda,\bar\rho,\bar\varphi,\bar\eta,\bar\chi,\bar{h},\bar\psi,\bar\xi,\bar{u}\}\,.
\end{equation}
Similarly, from the linearly broken parametric equations, eqs.\eqref{peqs}, it follows that  $\S^{count}$ is independent from
the mass parameters $\vartheta^{2}$, $m^{2}$, $M^{2}$, $\mu^{2}$ and $v^{2}$. Due to the nilpotency of the linearized Slavnov-Taylor operator $\mathcal{S}_{\Sigma}$, eq.\eqref{linst}, the characterization of the most general allowed counterterm corresponds to identifying the cohomology  \cite{Piguet:1995er} of the operator  $\mathcal{S}_{\Sigma}$ in the class of the integrated local polynomials of dimension four and zero ghost number, subject to the remaining constraints, eqs.\eqref{ct1}-\eqref{ct5}. After a lengthy algebraic analysis, it turns out that the most general invariant counterterm can be written as
\begin{equation}
\S^{count}=a_0\,S_{YM}[A]+a_0\,S_{YM}[B]+\mathcal{S}_{\S}\Delta^{(-1)}\label{count}
\end{equation}
with $\Delta^{(-1)}$ being given by
\begin{eqnarray}
\Delta^{(-1)}&=&\int d^{4}x\,\biggl(a_{1}\,\Omega^{a}_{\mu}A^{a}_{\mu}
+a_{1}\,\bar\Omega^{a}_{\mu}B^{a}_{\mu}
+a_{2}\,\bar{c}^{a}D^{ab}_{\mu}A^{b}_{\mu}
+a_{2}\,\bar{\omega}^{a}\bar{D}^{ab}_{\mu}B^{b}_{\mu}
+a_{3}\,c^{a}L^{a}
+a_{3}\,\omega^{a}\bar{L}^{a}\nonumber\\
&&-\frac{i\alpha}{2}\,a_4\,(\bar{c}^{a}b^{a}+ig\e^{ab}\bar{c}^{a}\bar{c}^{b}c)
-\frac{i\alpha}{2}\,a_4\,(\bar{\omega}^{a}\bar{b}^{a}+ig\e^{ab}\bar{\omega}^{a}\bar{\omega}^{b}\omega)\nonumber\\
&&+\frac{1}{2}\,(a_2+a_3+\alpha\,a_5)\,\varphi(A^{a}_{\mu}A^{a}_{\mu}+B^{a}_{\mu}B^{a}_{\mu})
-\alpha(a_3+a_4-a_6)\,\varphi(\bar{c}^{a}c^{a}+\bar{\omega}^{a}\omega^{a})\nonumber\\
&&+\frac{1}{2}\,(a_5\,\psi+a_7\,\lambda+a_9\,\chi)(A^{a}_{\mu}A^{a}_{\mu}+B^{a}_{\mu}B^{a}_{\mu})
+(a_6\,\psi+a_8\,\lambda+a_{10}\,\chi)(\bar{c}^{a}c^{a}+\bar{\omega}^{a}\omega^{a})\nonumber\\
&&+a_{11}\,g\e^{ab}u(\bar{c}^{a}c^{b}+\bar{\omega}^{a}\omega^{b})
+\frac{\zeta_{1}}{2}\,a_{12}\,\lambda\s
+\frac{\zeta_{2}}{2}\,a_{13}\,\varphi\rho
+\frac{\zeta_{3}}{2}\,a_{14}\,\psi{h}
+\frac{\zeta_{4}}{2}\,a_{15}\,\chi\eta
+\frac{\zeta_{5}}{2}\,a_{16}\,u\xi\nonumber\\
&&
+k_{1}\,a_{17}\,\lambda\rho
+k_{2}\,a_{18}\,\lambda\eta
+k_{3}\,a_{19}\,\lambda{h}
+k_{4}\,a_{20}\,\varphi\eta
+k_{5}\,a_{21}\,\varphi{h}
+k_{6}\,a_{22}\,\chi{h}\biggr)\,,  \label{a22}
\end{eqnarray}
and $a_n$, with $n=0,\dots,22$, being arbitrary dimensionless coefficients. \\\\It remains now to show that the counterterm \eqref{count} can be reabsorbed into the starting classical action $\Sigma$ through a suitable redefinition of the fields, sources and parameters. This will be the task of the next subsection.

%%%%%%%%%%%%%%%%%%%%%%%%%%%%%
\subsection{Determining the renormalization factors}
%%%%%%%%%%%%%%%%%%%%%%%%%%%%%
In order to show that the counterterm $\S^{count}$, eq.\eqref{count}, can be reabsorbed through a redefinition of the fields, sources and parameters, we shall employ the following compact notation. All fields will be denoted by $\{ f \}$, all sources by  $\{J \}$ and all parameters by  $\{p \}$.  As usual, the renormalization factors are introduced by
\begin{eqnarray}
f^{off}_{0}&=&\tilde{Z}_{f}^{1\!/2}\,f^{off}\,,\nonumber\\
f^{diag}_{0}&=&Z_{f}^{1\!/2}\,f^{diag}\,,\nonumber\\
J^{off}_{0}&=&\tilde{Z}_{J}\,J^{off}\,,\nonumber\\
J^{diag}_{0}&=&{Z}_{J}\,J^{diag}\,,\nonumber\\
p_{0}&=&Z_p\,p\,,
\end{eqnarray}
where the quantities labeled by "0" stand for the bare, or unrenormalized, quantities, while those with no label are the renormalized ones. Stating that the counterterm $\S^{count}$ can be reabsorbed into the starting action $\S$, amounts to prove that
\begin{equation}
\S[f_0,J_0,p_0]=\S[f,J,p]+\e\, \S^{count}[f,J,p] + O(\e^2)\,.  \label{reabs}
\end{equation}
Due to the mirror symmetry, it follows that
\begin{eqnarray}
&\tilde{Z}_{A}=\tilde{Z}_{U}\,,\quad
{Z}_{A}={Z}_{U}\,,\quad
\tilde{Z}_{b}=\tilde{Z}_{\bar{b}}\,,\quad
{Z}_{b}={Z}_{\bar{b}}\,,\quad
\tilde{Z}_{c}=\tilde{Z}_{\omega}\,,\quad
{Z}_{c}={Z}_{\omega}\,,&\nonumber\\
&\tilde{Z}_{\bar{c}}=\tilde{Z}_{\bar\omega}\,,\quad
{Z}_{\bar{c}}={Z}_{\bar\omega}\,,\quad
\tilde{Z}_{\Omega}=\tilde{Z}_{\bar\Omega}\,,\quad
{Z}_{\Omega}={Z}_{\bar\Omega}\,,\quad
\tilde{Z}_{L}=\tilde{Z}_{\bar{L}}\,,\quad
{Z}_{L}={Z}_{\bar{L}}\,.&
\end{eqnarray}
For what it concerns the renormalization factors of the linearizing auxiliary fields and of the mass parameters, we have to properly take into account that these variables can mix each other, due to the fact that they possess the same quantum numbers, {\it i.e.} the same dimensions, the same color structure , etc. As a consequence, the corresponding renormalization factors turns out to be expressed in terms of matrices, namely
\begin{equation}
\left(\begin{matrix}
\rho_0\cr h_0\cr\s_0\cr\eta_0
\end{matrix}\right)=\mathbb{Z}_{1}\,\left(\begin{matrix}
\rho\cr h\cr\s\cr\eta
\end{matrix}\right)\,,\quad
\left(\begin{matrix}
\bar\rho_0\cr \bar{h}_0\cr\bar\s_0\cr\bar\eta_0
\end{matrix}\right)=\mathbb{Z}_{2}\,\left(\begin{matrix}
\bar\rho\cr \bar{h}\cr\bar\s\cr\bar\eta
\end{matrix}\right)\,,\quad
\left(\begin{matrix}
\varphi_0\cr \psi_0\cr\lambda_0\cr\chi_0
\end{matrix}\right)=\mathbb{Z}_{3}\,\left(\begin{matrix}
\varphi\cr\psi\cr\lambda\cr\chi
\end{matrix}\right)\,,\quad
\left(\begin{matrix}
\bar\varphi_0\cr \bar{\psi}_0\cr\bar\lambda_0\cr\bar\chi_0
\end{matrix}\right)=\mathbb{Z}_{4}\,\left(\begin{matrix}
\bar\varphi\cr \bar{\psi}\cr\bar\lambda\cr\bar\chi
\end{matrix}\right)\,,
\end{equation}
\begin{equation}
\left(\begin{matrix}
m^2_0\cr M^2_0\cr\vartheta^2_0\cr\mu^2_0
\end{matrix}\right)=\mathbb{Z}_{5}\,\left(\begin{matrix}
m^2\cr M^2\cr\vartheta^2\cr\mu^2
\end{matrix}\right)\,.
\end{equation}
All the renormalization factors, even the $\mathbb{Z}$ matrices, coincide with  unity at zeroth  order in the perturbative expansion,
\begin{eqnarray}
Z&=&1+O(\e)\,,\nonumber\\
\mathbb{Z}&=&\mbox{\bf 1}+O(\e)\,.
\end{eqnarray}
After direct inspection, it turns out that condition \eqref{reabs} can be consistently fulfilled with the $Z$'s factors determined in the following way:
\begin{eqnarray}
\tilde{Z}^{1\!/2}_{A}&=&1+\e\,\Bigl(\frac{a_{0}}{2}+a_{1}\Bigr)\,,\nonumber\\
Z_{g}&=&1-\e\,\frac{a_{0}}{2}\,,\nonumber\\
\tilde{Z}^{1\!/2}_{c}&=&1+\e\,\Bigl(\frac{a_2}{2}-\frac{a_3}{2}\Bigr)\,,\nonumber\\
{Z}^{1\!/2}_{c}&=&1+\e\,\Bigl(\frac{a_2}{2}+\frac{a_3}{2}\Bigr)\,,\nonumber\\
{Z}^{1\!/2}_{\xi}&=&1+\e\,\Bigl(\frac{a_0}{2}-a_2+a_{11}\Bigr)\,,\nonumber\\
Z_{\alpha}&=&1+\e\,\bigl(a_0-2a_2+a_4\bigr)\,,\nonumber\\
Z_{\zeta_{1}}&=&1+\e\,\bigl(a_{12}+2a_{0}\bigr)\,,\nonumber\\
Z_{\zeta_{2}}&=&1+\e\,\bigl(a_{13}+2a_{0}-2a_{3}-2\alpha a_{5}\bigr)\,,\nonumber\\
Z_{\zeta_{3}}&=&1+\e\,\bigl(a_{14}+2a_{2}-2a_{6}\bigr)\,,\nonumber\\
Z_{\zeta_{4}}&=&1+\e\,\bigl(a_{15}-2a_{0}\bigr)\,,\nonumber\\
Z_{\zeta_{5}}&=&1+\e\,\bigl(a_{16}-a_{0}+2a_{2}-2a_{11}\bigr)\,,\nonumber\\
Z_{k_{1}}&=&1+\e\,\Bigl(a_{17}-a_2-a_3-\alpha a_5-\frac{2\zeta_{2}}{k_{1}}\,a_7\Bigl)\,,\nonumber\\
Z_{k_{2}}&=&1+\e\,\bigl(a_{18}-2a_0\bigr)\,,\nonumber\\
Z_{k_{3}}&=&1+\e\,\Bigl(a_{19}-a_0+a_2-a_6-\frac{2\zeta_{3}}{k_{3}}\,a_8\Bigl)\,,\nonumber\\
Z_{k_{4}}&=&1+\e\,\Bigl(a_{20}-a_2-a_3-\alpha a_5-\frac{2\zeta_{2}}{k_{4}}\,a_9\Bigl)\,,\nonumber\\
Z_{k_{5}}&=&1+\e\,\Bigl[a_{21}+a_0-\Bigl(1-\frac{\alpha\zeta_{3}}{k_{5}}\Bigr)\,a_3
+\frac{\alpha\zeta_{3}}{k_{5}}\,a_4-\Bigl(\alpha-\frac{2\zeta_{2}}{k_{5}}\Bigr)\,a_5
-\Bigl(1+\frac{\alpha\zeta_{3}}{k_{5}}\Bigr)\,a_6\Bigl]\,,\nonumber\\
Z_{k_{6}}&=&1+\e\,\Bigl(a_{22}-a_0+a_2-a_6-\frac{2\zeta_{3}}{k_{6}}\,a_{10}\Bigl)\,,\label{Zs}
\end{eqnarray}
and
\begin{equation}
\mathbb{Z}_{1}=\mathbf{1}+\e
\left(\begin{matrix}
-a_0+a_2+a_3+\alpha a_5&\Bigl|&a_5&\Bigl|&a_7&\Bigl|&a_9\cr
-\alpha(a_3+a_4-a_6)&\Bigl|&-a_2+a_6&\Bigl|&a_8&\Bigl|&a_{10}\cr
0&\Bigl|&0&\Bigl|&a_0&\Bigl|&0\cr
0&\Bigl|&0&\Bigl|&0&\Bigl|&a_0
\end{matrix}\right)\,,
\end{equation}

\begin{equation}
\mathbb{Z}_{2}=\mathbf{1}+\e
\left(\begin{matrix}
a_0-a_2-a_3-\alpha a_5&\Bigl|&\alpha(a_3+a_4-a_6)&\Bigl|&0&\Bigl|&0\cr
-a_5&\Bigl|&a_2-a_6&\Bigl|&0&\Bigl|&0\cr
-a_7&\Bigl|&-a_8&\Bigl|&-a_0&\Bigl|&0\cr
-a_9&\Bigl|&-a_{10}&\Bigl|&0&\Bigl|&-a_0
\end{matrix}\right)\,,
\end{equation}

\begin{equation}
\mathbb{Z}_{3}=\mathbf{1}+\e\!
\left(\begin{matrix}
-\frac{1}{2}(a_0-a_2-a_3)+\alpha a_5\!\!&\!\!\Bigl|\!\!&\!\!a_5\!\!&\!\!\Bigl|\!\!&\!\!a_7\!\!&\!\!\Bigl|\!\!&\!\!a_9\cr
-\alpha(a_3+a_4-a_6)\!\!&\!\!\Bigl|\!\!&\!\!\frac{1}{2}(a_0-3a_2-a_3)+a_6\!\!&\!\!\Bigl|\!\!&\!\!a_8\!\!
&\!\!\Bigl|\!\!&\!\!a_{10}\cr
0\!\!&\!\!\Bigl|\!\!&\!\!0\!\!&\!\!\Bigl|\!\!&\!\!2a_7-a_0-a_2-a_3\!\!&\!\!\Bigl|\!\!&\!\!0\cr
0\!\!&\!\!\Bigl|\!\!&\!\!0\!\!&\!\!\Bigl|\!\!&\!\!0\!\!&\!\!\Bigl|\!\!&\!\!-\frac{1}{2}(a_0+a_2+a_3)
\end{matrix}\right)\,,
\end{equation}

\begin{equation}
\mathbb{Z}_{4}=\mathbf{1}+\e\!
\left(\begin{matrix}
\frac{1}{2}(a_0-a_2-a_3)-\alpha a_5\!\!&\!\!\Bigl|\!\!&\!\!\alpha(a_3+a_4-a_6)\!\!&\!\!\Bigl|\!\!&\!\!0\!\!&\!\!\Bigl|\!\!&\!\!0\cr
-a_5\!\!&\!\!\Bigl|\!\!&\!\!-\frac{1}{2}(a_0-3a_2-a_3)-a_6\!\!&\!\!\Bigl|\!\!&\!\!0\!\!
&\!\!\Bigl|\!\!&\!\!0\cr
-a_7\!\!&\!\!\Bigl|\!\!&\!\!-a_8\!\!&\!\!\Bigl|\!\!&\!\!-2a_7+a_0+a_2+a_3\!\!&\!\!\Bigl|\!\!&\!\!0\cr
-a_9\!\!&\!\!\Bigl|\!\!&\!\!-a_{10}\!\!&\!\!\Bigl|\!\!&\!\!0\!\!&\!\!\Bigl|\!\!&\!\!\frac{1}{2}(a_0+a_2+a_3)
\end{matrix}\right)\,,
\end{equation}

\begin{equation}
\mathbb{Z}_{5}=\mathbf{1}+\e
\left(\begin{matrix}
-a_0+a_2+a_3+\alpha a_5&\Bigl|&a_5&\Bigl|&i\sqrt{2}\,a_7&\Bigl|&a_9\cr
-\alpha(a_3+a_4-a_6)&\Bigl|&-a_2+a_6&\Bigl|&i\sqrt{2}\,a_8&\Bigl|&a_{10}\cr
0&\Bigl|&0&\Bigl|&a_0&\Bigl|&0\cr
0&\Bigl|&0&\Bigl|&0&\Bigl|&a_0
\end{matrix}\right)\,.
\end{equation}
The remaining renormalizations factors are not independent, being given by
\begin{eqnarray}
&Z^{1\!/2}_{A}=Z^{-1}_{g}\,,\qquad\tilde{Z}^{1\!/2}_{\bar{c}}=\tilde{Z}^{1\!/2}_{{c}}\,,\qquad
{Z}^{1\!/2}_{\bar{c}}={Z}^{-1\!/2}_{{c}}\,,\qquad
\tilde{Z}^{1\!/2}_{b}=Z_g{Z}^{1\!/2}_{{c}}\tilde{Z}^{1\!/2}_{{c}}\,,\qquad
{Z}^{1\!/2}_{b}={Z}^{-1}_{g}\,,&\nonumber\\
&\tilde{Z}_{\Omega}=Z_g^{-1}\tilde{Z}^{-1\!/2}_{A}{Z}^{-1\!/2}_{{c}}\,,\qquad
Z_{\Omega}=Z_{c}^{1\!/2}\,,\qquad
\tilde{Z}_{L}=Z_g^{-1}{Z}^{-1\!/2}_{c}\tilde{Z}^{-1\!/2}_{{c}}\,,\qquad
{Z}_{L}=Z_g^{-1}{Z}^{-1}_{{c}}\,,&\nonumber\\
&Z^{1\!/2}_{\bar\xi}=Z^{-1\!/2}_{\xi}\,,\qquad
Z_{v^{2}}=Z^{1\!/2}_{\xi}\,,\qquad
Z^{1\!/2}_{u}=Z_{g}^{-1}Z_{c}^{1\!/2}Z^{1\!/2}_{\xi}\,,\qquad
Z^{1\!/2}_{\bar{u}}=Z_{g}Z_{c}^{-1\!/2}Z^{-1\!/2}_{\xi}\,.&
\end{eqnarray}
This proves that the action $\S$ of the replica model in the maximal Abelian gauge is renormalizable.

%%%%%%%%%%%%%%%%%%%%%%%%%%%%%%%%%%%%%%%%%%%%%%%%%%%%%%%%%%%%%%%%%%%%%%%%%%%%%%%%%%%%%%%%%%%%%%%%%%%%%%%%%%%%%%%%%

\end{document}